\documentclass[aps,onecolumn,preprint,superscriptaddress,nofootinbib,floats]{revtex4}
\usepackage{amsmath,amssymb,color,mathrsfs, graphicx,verbatim,epsfig, bbm, wasysym}
\usepackage[hyperfootnotes=false]{hyperref}
\usepackage{slashed}
\allowdisplaybreaks

\usepackage{graphicx}
\usepackage{url}
\usepackage{color}
\usepackage{amsmath}
\usepackage{amsfonts}
\usepackage{amssymb}
\def\beq{\begin{equation}}
\def\eeq{\end{equation}}
\def\beqa{\begin{eqnarray}}
\def\eeqa{\end{eqnarray}}

\def\ltap{\ \raise.3ex\hbox{$<$\kern-.75em\lower1ex\hbox{$\sim$}}\ }
\def\gtap{\ \raise.3ex\hbox{$>$\kern-.75em\lower1ex\hbox{$\sim$}}\ }
\newcommand{\missET}{\slash{\hspace{-2.5mm}E}_T}

\def\stop{\tilde t}
\def\mstop{m_{\tilde t}}

\begin{document}

\baselineskip 0.6cm

\begin{titlepage}

\thispagestyle{empty}

\begin{flushright}
PITT PACC 1309
\end{flushright}

\begin{center}

\vskip 0.7cm

{\Large \bf Pulling Out All the Stops: \\ \vspace{0.1cm} Searching for RPV SUSY with Stop-Jets}

\vskip 1.0cm
{\large  Yang Bai$^{a}$, Andrey Katz$^{b}$,  and Brock Tweedie$^{c,d}$}
\vskip 0.4cm
{\it $^a$ Department of Physics, University of Wisconsin, Madison, WI 53706} \\
{\it $^b$ Center for the Fundamental Laws of Nature, Jefferson Physical Laboratory, \\Harvard University, Cambridge, MA 02138} \\
{\it $^c$ Physics Department, Boston University, Boston, MA 02215} \\
{\it $^d$ PITT PACC, Department of Physics and Astronomy, University of Pittsburgh, Pittsburgh, PA 15260}
\vskip 1.2cm

\end{center}

\noindent  If the lighter stop eigenstate decays directly to two jets via baryonic R-parity violation, it could have escaped existing LHC and 
Tevatron searches in four-jet events, even for masses as small as 100~GeV.  
In order to recover sensitivity in the face of increasingly harsh trigger requirements at the LHC, we propose a search for stop pairs in 
the highly-boosted regime, using the approaches of jet substructure.  We demonstrate that the four-jet triggers can be completely 
bypassed by using inclusive jet-$H_T$ triggers, and that the resulting QCD continuum background can be processed by substructure methods into 
a featureless spectrum suitable for a data-driven bump-hunt down to 100~GeV.  We estimate that the 
LHC~8~TeV run is sensitive to 100~GeV stops with decays of any flavor at better than 5$\sigma$-level, and could place exclusions up to 300~GeV or higher.  
Assuming Minimal Flavor Violation and running a $b$-tagged analysis, exclusion reach may extend up to nearly 400~GeV.  Longer-term, the 14~TeV LHC at 300~fb$^{-1}$ could extend these 
mass limits by a factor of two, while continuing to improve sensitivity in the 100~GeV region.

\end{titlepage}

\setcounter{page}{1}

\section{Introduction}
\label{sec:intro}


Well over one hundred null searches for supersymmetric signals at the LHC have now been completed, leaving us to ponder the fate of naturalness after the Higgs boson discovery.  As we await the energy upgrade, much attention is being focused on the possibility that supersymmetric particles {\it have} been produced at the LHC, but for one reason or another are buried amidst the copious Standard Model backgrounds.  In this paper, we turn to one of the simplest of these possibilities:  the lightest superpartner is a stop eigenstate, and decays promptly to two quarks through a baryonic R-parity violating (RPV) coupling.

Such a situation is in fact quite well-motivated within the context of ``effective'' or ``natural'' 
supersymmetry~\cite{Dimopoulos:1995mi, Cohen:1996vb, Kats:2011qh, Brust:2011tb, Papucci:2011wy}, 
where only the superparticles required to regulate the Higgs mass need to be within the immediate reach of the LHC, and the detailed spectrum and dynamics above the multi-TeV scale can be left unspecified.  In this framework, third generation squarks can be amongst the lightest superparticles, forcing us to take seriously the possibility that a stop eigenstate sits at or near the bottom of the SUSY spectrum.  R-parity conservation is still often assumed in natural SUSY, but its violation must also be considered.  The main motivations for R-parity are stabilization of a supersymmetric dark matter candidate and suppression of proton decay.  The dark matter motivation is attractive, but the issues of dark matter and electroweak naturalness might easily be decoupled.  The proton decay motivation does not require R-parity per se, but only that baryon-number violation (BNV) and lepton-number violation (LNV) are not simultaneously active.  Moreover, the motivation for R-parity is further weakened by the presence of R-parity-conserving operators at dimension-5 with simultaneous BNV and LNV.  In a relatively model-agnostic approach with a multi-TeV cutoff, these operators in any case force us to consider proton-stabilizing symmetries other than R-parity.  RPV is also sometimes considered dangerous given precision tests, in particular $K-\bar K$ and $n - \bar n$ mixing in the BNV case.  However, couplings that are small enough to avoid these constraints are still usually large enough to allow for prompt RPV 
decays from the perspective of the LHC~\cite{Brust:2012uf} (see~\cite{Barbier:2004ez} for a complete list of constraints).

\begin{figure}[ht!]
\begin{center}
\includegraphics[width=0.70\textwidth]{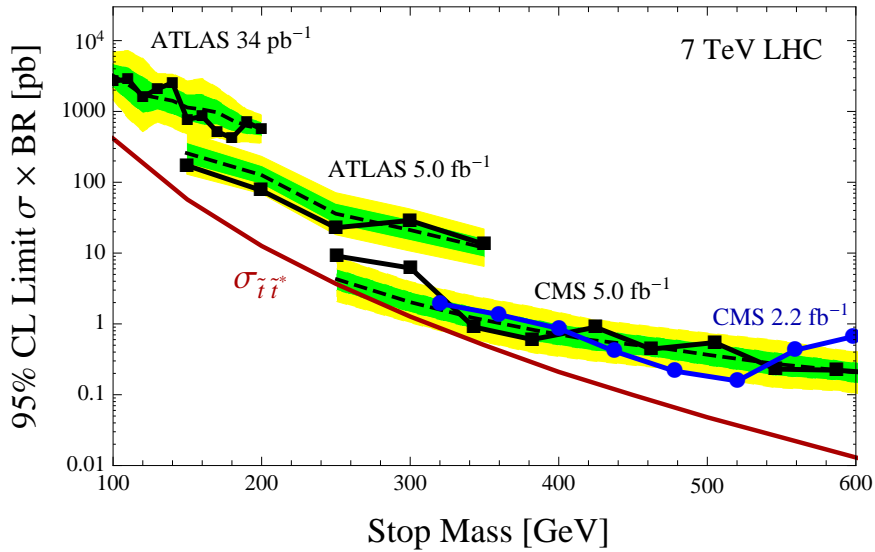}
\caption{Existing constraints on $pp \to \tilde t \tilde t^* \to 4j$ from the LHC, reinterpreting the results of~\cite{CMS:coloron,Chatrchyan:2013izb,Aad:2011yh,Atlas:coloron} to account for stop acceptances relative to coloron or hyperpion acceptances.}
\label{fig:LHCconstraints}
\end{center}
\end{figure}

The idea that SUSY might be both ``natural'' and violate R-parity has received increasing attention in the past few years~\cite{Brust:2011tb,Evans:2012bf,Csaki:2011ge,Graham:2012th,Krnjaic:2012aj,Bhattacherjee:2013gr,Franceschini:2013ne,Csaki:2013we,Florez:2013mxa,Krnjaic:2013eta,Monteux:2013mna,DiLuzio:2013ysa,Csaki:2013jza}, 
exactly because the relatively small production cross sections and non-canonical decay topologies lend themselves to evading LHC searches.  
Many spectra and choices of active RPV operators exist, and the nontrivial flavor structures of the couplings must be specified.  
In some cases, especially those involving leptonic RPV, exclusion or discovery prospects are quite good (see~\cite{Evans:2012bf} for proposals and recast limits, and e.g.~\cite{Chatrchyan:2013xsw,ATLAS-2013-036} for some recent experimental searches).  Successful baryonic RPV searches have also been performed, such as LSP gluinos decaying fully hadronically into multiple jets~\cite{ATLAS-2013-091,ATLAS:2012dp,CMS-PAS-EXO-12-049,Chatrchyan:2012uxa}.  However, an LSP stop decaying promptly to two jets via baryonic RPV has proven particularly evasive.  Stop pairs are produced at much smaller rates than gluino pairs of comparable mass, and the lower-multiplicity 4-jet final state is even more difficult to disentangle from the pure QCD backgrounds.  Another major complicating aspect at the LHC is the multijet triggers, 
which can heavily prescale-away the signatures of stops lighter than several hundred GeV.  Some of the best current direct limits actually come from LEP, 
which rules out $\mstop \ltap 90$~GeV~\cite{Heister:2002jc}.  A recent search at the Tevatron extends this limit up to only about 100~GeV~\cite{Aaltonen:2013hya}.  
However, so far, direct searches for pair-production of dijet resonances at the LHC have failed to reach the sensitivity necessary to place constraints for any stop 
mass~\cite{CMS:coloron,Chatrchyan:2013izb,Aad:2011yh,Atlas:coloron}.  A snapshot of the current situation can be seen in Fig.~\ref{fig:LHCconstraints}.  In fact, the inevitable rise of trigger thresholds with instantaneous luminosity and beam energy leaves us to wonder whether the LHC will {\it ever} be sensitive to this signal.  At the very least, this trend suggests that masses near the current limit of 100~GeV might be left unexplored.\footnote{For recent projections for the long-term LHC, which begin to achieve exclusion reach but nonetheless do not pursue signals below 300~GeV, see the recent Snowmass study~\cite{Duggan:2013yna}.}

One way around these difficulties is to search for the stop as a dijet resonance produced in the decays of heavier colored superparticles, such as 
gluinos~\cite{Han:2012cu} or 
sbottoms~\cite{Brust:2012uf} (or possibly the heavier stop eigenstate), or to simply set bounds using the associated leptonic activity and high $H_T$ of these decays~\cite{Lisanti:2011tm,Allanach:2012vj,Berger:2013sir,ATLAS-2013-007}.  Naturalness suggests that these colored superparticles should also not be far above 1~TeV, and might be produced with observable rates.  It is also possible to invoke Minimal Flavor Violation (MFV), which suggests that stops dominantly decay (with a branching ratio$\simeq 95\%$) into $\bar b \bar s$ or $\bar b \bar d$~\cite{Csaki:2011ge}.  
It was pointed out in~\cite{Franceschini:2012za} that incorporating $b$-tagging into the triggering might allow the direct stop pair signal to write to tape with higher efficiency, and subsequent kinematic analysis can discriminate it from flavored backgrounds.  The use of $b$-tagging is also usually considered to help identify stop production in gluino decay.  However, at present, there seems to be no strategy at the LHC that guarantees full coverage of the general $\stop\stop^* \to 4j$ signal, exploiting neither spectrum-dependent nor flavor-dependent features of the theory.

Rather than consider the generic signal lost, we here exploit what is now a well-worn trick~\cite{Seymour:1993mx,Butterworth:2002tt,Butterworth:2008iy}:  we focus on ``boosted'' production at high transverse momentum, and apply the methods of jet substructure to ``stop-jets," each containing an entire stop decay.  Indeed, an analysis of this type has already been applied in the search for light RPV gluinos~\cite{ATLAS:2012dp}, and has been studied for color-octet scalars~\cite{Bai:2011mr}.  The advantages of a boosted/substructure analysis are manyfold, including better $S/B$, automatic resolution of combinatoric ambiguities, and improved mass resolution due to more complete decay radiation containment and rejection of uncorrelated soft radiation.  But probably the primary advantage for a stop pair search is in how a substructure treatment can process the continuum QCD background.  The standard resolved 4-jet analyses partition the leading jets into two pairs so as to form two stop candidates, either minimizing the candidate mass asymmetry (CMS) or minimizing a $\Delta R$ measure for the two decays (ATLAS).  The stop is searched for as a bump in the average pair-mass spectrum.  In either case, the requirement of four well-separated and high-$p_T$ jets passing the triggers, as well as other downstream analysis cuts, leads to a highly-shaped background QCD spectrum with a pronounced ``trigger turn-on'' peaking at 100--200~GeV.  As a consequence, the most recent CMS~\cite{Chatrchyan:2013izb} (ATLAS~\cite{Atlas:coloron}) analysis of 2011 data does not even search for masses below 250~GeV (150~GeV).

By contrast, as we will see, a carefully constructed jet substructure approach potentially leads to a largely featureless average-mass spectrum for the QCD background from a few 10's of GeV up.  This is because jet substructure, unlike traditional jet analyses, gives us far more flexibility in assigning sprays of hadrons to individual hard ``quarks'' or ``gluons.''  In particular, we can make these assignments in a way much closer to QCD itself, which is approximately a scale-invariant theory.  To best exploit this feature, we will also capitalize on the summed jet-$H_T$
trigger rather than triggers that count specific numbers of jets above some threshold.  
The jet-$H_T$ trigger, which exists in some form in both CMS and ATLAS, probably has the least sensitivity to precisely how the event activity groups into standard LHC jets.

With an unbiased background spectrum, a bump-hunt becomes much more tractable, regardless of what stop mass we consider.  Either a parametrized spectral fit or kinematic sideband approaches like the ABCD method can be employed to determine the QCD background, and we explore the viability of several such approaches, including some novel ones.  We find that 100~GeV stops should be visible in the 2012 data set with better than 5$\sigma$ statistical significance.  Exclusion reach should extend up to more than 300~GeV.  We also consider what might be possible in the MFV case, using a $b$-tagged version of the analysis.  $S/B$ improves dramatically, for example yielding 10$\sigma$-level statistical sensitivity to 100~GeV stops, and exclusion up to almost 400~GeV.  For the longer-term Run II of the LHC, with a projected 300~fb$^{-1}$ at 14~TeV, we estimate untagged exclusions extending up to 650~GeV, with discovery-level sensitivity or better between 100~GeV and 500~GeV.

Our paper is organized as follows.  In the next section, we discuss our substructure procedures, and explore the response of the signal and QCD continuum in simulation data.  We present our data analysis techniques and final sensitivity estimates in Section~\ref{sec:discovery}.  We conclude in Section~\ref{sec:conclusions}.  Two appendices contain more details of our simulations and more in-depth substructure studies.


\section{Jet Substructure Techniques and Analysis Cuts}
\label{sec:techniques}


The fundamental obstacle to any multijet search at the LHC is the overwhelming production rate for continuum QCD backgrounds.  The dynamics of QCD is approximately scale-invariant at high energy.  Convolved with smoothly-falling parton distribution functions, a good stage is set for extracting signals with sharply-localized mass features, such as RPV stop pairs.  However, the standard approaches to reconstructing QCD events are far from scale-invariant.  ``Jets'' as usually defined are tied to a dimensionful $p_T$ threshold, as well as a dimensionless jet radius $R$.  For example, in a multijet search with $p_T(j) > 100$~GeV and $R=0.5$, there is an absolute minimum invariant mass between jet pairs of approximately 50~GeV.  Any QCD background invariant mass spectrum constructed from such jets will start at 50~GeV, rise at higher invariant masses as the efficiency turns on, and only then turn over into a smoothly-falling shape.  Searches for low-$S/B$ features in the broad turn-on region or at the peak are often not considered, as the precise signal and background shapes in these regions have high sensitivity to reconstruction uncertainties.

Indeed, these kinds of considerations have limited the range of applicability of current LHC stop searches based on the multijet approach, as seen in Fig.~\ref{fig:LHCconstraints}.  To control triggering rates as the LHC continues to increase instantaneous luminosity and energy, the jet $p_T$ thresholds are gradually increasing.  For example, to keep 4-jet triggering rates at their 2012 levels during the projected 300~fb$^{-1}$ run at 14~TeV, offline $p_T(j)$ thresholds would need to be increased to 150--200~GeV.  Stops sitting near the current exclusion limit of $m = 100$~GeV might therefore never be visible.  Even with 2011 data, only the very low-luminosity ATLAS search~\cite{Aad:2011yh} was capable of probing masses near that limit, but was not sensitive to stop pair cross sections.  Despite the fact that the cross section is becoming quite large near 100~GeV, basic triggering and analysis cuts at nominal LHC luminosities are carving away the signal.

As has been pointed out in~\cite{Katz:2010mr,Gouzevitch:2013qca}, reconstruction-induced biases can be highly ameliorated by applying modern approaches to jet substructure on ``fat-jets.''  Jet substructure can be less tied to fixed $p_T$ or $\Delta R$ thresholds for the individual reconstructed hard partons, and can put more focus on dimensionless quantities such as ratios of masses or $p_T$'s.  The ``price'' for applying a substructure-based search is that we must work in the boosted region of production phase space, which might still represent only a small fraction of the total signal.  Nonetheless, as we will see, the tradeoff is more than worthwhile.  In any case, existing 4-jet searches are already capitalizing on the boosted or semi-boosted region of phase space, which offers the additional advantages of better $S/B$ and much-reduced combinatoric ambiguities.  

We are in the midst of an ongoing boom of jet substructure ideas, many of which can have applicability to the boosted stop pair signal.  We here focus on the relative-$p_T$ declustering method used in the JHU top-tagger~\cite{Kaplan:2008ie} and the diboson-jet tagger of~\cite{Son:2012mb}, which is itself based on the 
``BDRS'' method of~\cite{Butterworth:2008iy}.   
These methods identify localized clusters of energy within the fat-jets as ``subjets,'' and subsequently treat these objects similar to ordinary QCD jets.  However, before proceeding, we emphasize that ideas such as (but not limited to) pruning~\cite{Ellis:2009su}, N-subjettiness~\cite{Thaler:2010tr}, template overlaps~\cite{Almeida:2010pa}, shower deconstruction~\cite{Soper:2011cr}, and Q-jets~\cite{Ellis:2012sn} could all  worth more dedicated study in this context, and might lead to further improvements.

For our fat-jet clustering radius, we choose a very large value of $R = 1.5$, roughly giving us hemisphere-sized jets.  The motivation for such a large radius is twofold.  First, the large catchment area gives us a broader reach in stop masses.  For given momentum thresholds, heavier stops will be able to pass with less boost, and therefore will have more widely-separated decay products.  Heavier stops also have much smaller cross section, and the very highly-boosted portion of phase space may be too poorly-populated to be exploited.  In fact, for much of the mass range probed in this paper, the quarks in the stop decays are well-separated enough to be reconstructed as individual jets of normal radius.  However, we find it interesting that these masses can alternatively be covered using substructure procedures, raising the question of whether a traditional analysis is even necessary.  The second motivation for the large radius is more subtle, and has to do with how the continuum QCD background is processed.  Interplaying with momentum and substructure cuts, the jet radius sets a maximum mass (analogous to the {\it minimum} mass for normal jet-pair masses described above).  Larger radii push the turn-off of the mass spectrum out to higher values, creating a less steeply-falling background.  Larger radii also somewhat improve $S/B$.  This is because enlarging the jet radius increases sensitivity to wider-angle radiation, and provides a primitive form of color discrimination.  At high momenta, stop pair production is dominated by $q\bar q \to \tilde t \tilde t^*$, whereas the QCD background contains more highly-colored processes such as $qg \to qggg$.  (For a comparison of different $R$'s, see Appendix~\ref{sec:substructure}.)

Fixing $R=1.5$, we cluster jets in each event using the Cambridge/Aachen (C/A) algorithm~\cite{Dokshitzer:1997in,Wobisch:1998wt} as implemented in {\tt FastJet3}~\cite{Cacciari:2005hq}, and study the leading two jets within $|\eta| < 2.5$.  Then we iteratively undo the clustering stages individually within each jet.  Now viewed in reverse as a splitting, at each stage we can look at the two branches ``$a$'' and ``$b$'' and decide whether the splitting should be considered ``hard'' or ``soft'' according to some prescription.  For example, a soft splitting might resolve a low-$p_T$ collection of radiation at the jet's periphery.  
In our specific implementation, derived directly from the first declustering stage of the diboson-jet-tagger of Ref.~\cite{Son:2012mb}, a splitting is considered hard if both $a$ and $b$ carry appreciable $p_T$ relative to the original fat-jet, and if their individual $m/p_T$ ratios are not too large.  Explicitly, $\min[p_T(a),p_T(b)]/p_T({\rm fat\mbox{-}jet}) > 0.1$, $m(a)/p_T(a) < 0.3$ and $m(b)/p_T(b) < 0.3$. 
Otherwise the splitting is considered soft, in which case the lower-$p_T$ branch is thrown away, and the declustering is continued along the surviving branch.  The procedure is repeated until a hard splitting is encountered, or no more jet constituents remain (in which case the entire event is vetoed).  The two branches at the hard splitting are our two subjets, representing our assignment of the jet's radiation to the two quarks in the stop decay.\footnote{The original BDRS method uses a very similar procedure~\cite{Butterworth:2008iy}, relying instead on a mass-drop criterion and a somewhat different momentum-asymmetry criterion.  While BDRS can be adapted for use in the boosted stop search, there are important caveats which we discuss in more detail in Appendix~\ref{sec:substructure}.}  The reconstructed stop candidate is the four-vector-sum of these two subjets.

Most of our ability to discriminate stop pair production from ordinary QCD stems from the fact that the stop events contain two subjet-pair resonances of equal mass.  Therefore, instead of looking for a bump within the distribution of the individual stop candidate masses $m_1$ and $m_2$, we look for a bump in the joint distribution of $(m_1,m_2)$.  Practically, we can turn this into a 1D bump-hunt by first focusing on the region of small mass asymmetry $A \equiv |m_1-m_2|/(m_1+m_2)$, and then constructing the spectrum of the averaged mass $m_{\rm avg} \equiv (m_1+m_2)/2$.  We pick a nominal mass asymmetry threshold of 10\%. 

We have found that $S/B$ can be further purified by a handful of additional cuts.  In particular, we can place a cut on the stop-pair CM-frame production angle (as is done in~\cite{Aad:2011yh,Atlas:coloron}), formed by actively boosting the entire 4-subjet system to rest in the lab frame and measuring the angle of either stop candidate with respect to the beamline.  We call this angle $\theta^*$, and place a cut $|\cos\theta^*| < 0.3$.  Finally, we exploit the fact that energy splittings within QCD jets tend to be very asymmetric, whereas in stop decays they tend to be more democratic.  Within each stop pair, we demand that the $p_T$'s of subjets $a$ and $b$ relative to one another satisfy $\min[p_T(a),p_T(b)]/\max[p_T(a),p_T(b)] > 0.3$.

The final ingredient needed to define an analysis is to establish a trigger for the events.  We choose the total jet-$H_T$ trigger of CMS, which sums up the $p_T$'s of all ordinary jets.\footnote{ATLAS also has a dedicated single-fat-jet trigger, though this is defined with $R=1.0$.  Nonetheless, this trigger could still be used as the basis of a substructure analysis very similar to the one that we explore here.}  In Ref.~\cite{Chatrchyan:2012jx}, based on 2011 data, the $H_T$ trigger was used for a classic jets+$\missET$ style SUSY search.  The trigger was found to be fully efficient for final reconstructed events with $H_T > 750$~GeV, summing over $R=0.5$ anti-$k_T$ jets with $p_T > 50$~GeV.  Anticipating slightly harsher triggers in 2012, we conservatively set our threshold at 900~GeV.\footnote{A looser offline $H_T$ cut may in fact be feasible with 2012 data, and would only improve our results.  We thank Keith Ulmer for bringing this point to our attention.}

We test our substructure methods and cuts on signal and QCD background simulation samples for the 8~TeV LHC.  Full details of their generation can be found in Appendix~\ref{sec:simulations}.  We note here that the signal samples are matched up to one jet emission at the production stage, which gives a more accurate $H_T$ spectrum.  The QCD continuum simulations are matched up to four hard partons (including $b$-quarks), using the CKKW-L~\cite{Lonnblad:2001iq} prescription implemented in {\tt Pythia8}~\cite{Lonnblad:2011xx}.  We take a quark-flavor-conscious Durham $k_T$ distance as our merging measure, and a merging threshold of 50~GeV.  Other background samples, such as $t\bar t$, are generated in {\tt MadGraph5} interfaced with {\tt Pythia8}, or self-contained within {\tt Pythia8}.  All samples are processed through a simple, perfect $0.1 \times 0.1$ grid ``calorimeter'' in $\eta$--$\phi$ space.  To prove the robustness of our methods in a high pileup environment, we also introduce pileup events into the simulations and then apply a form of event-wide trimming~\cite{Krohn:2009th} tailored to pileup removal before the fat-jet clustering stage.  (This procedure also heavily reduces the impact of the underlying event and soft ISR.)  After declustering the fat-jets, individual subjets are energy-smeared similar to normal LHC jets (see Appendix~\ref{sec:simulations} for further details).

\begin{figure*}[ht!]
\begin{center}
\includegraphics[width=0.44\textwidth]{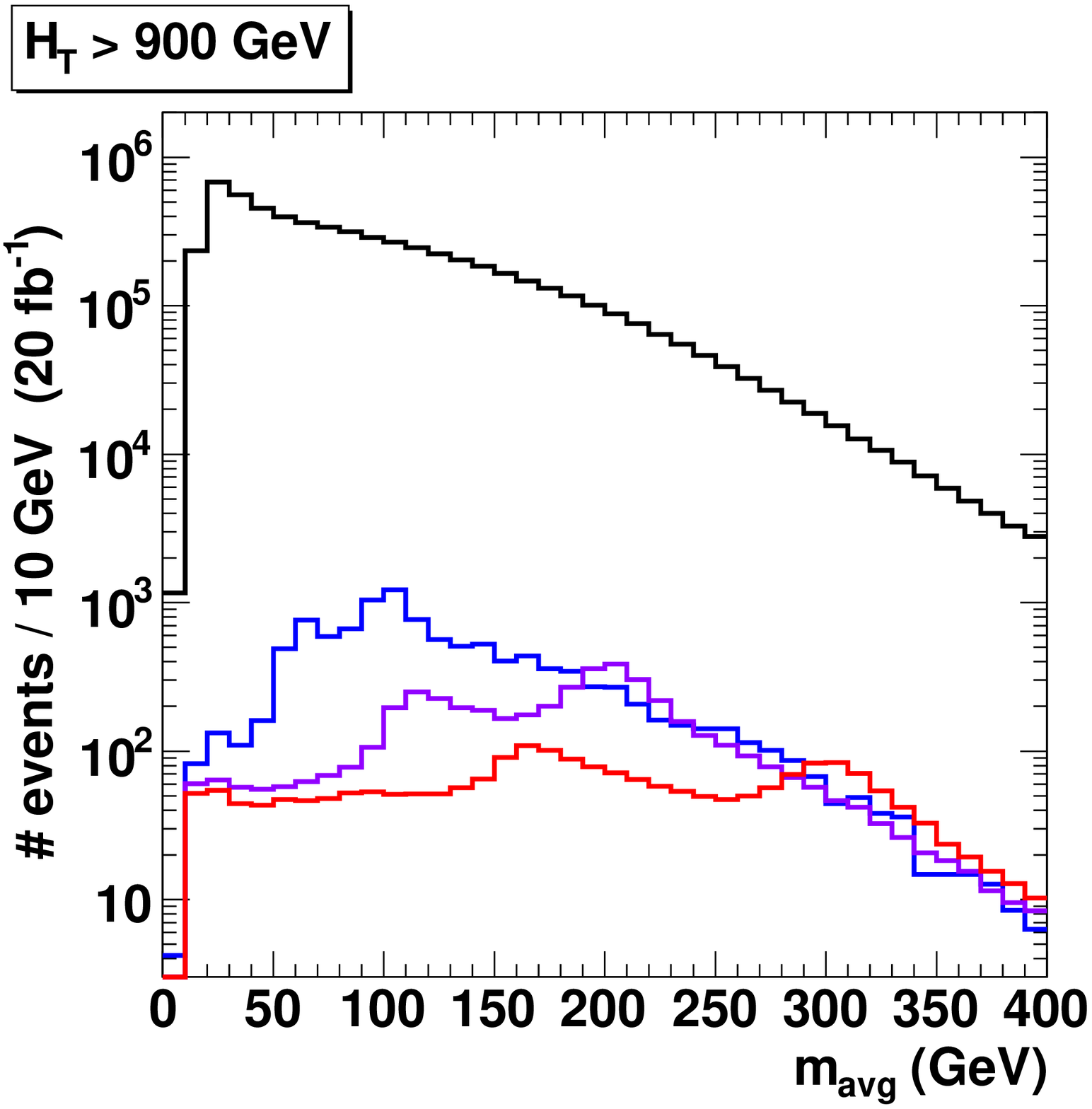}
\includegraphics[width=0.44\textwidth]{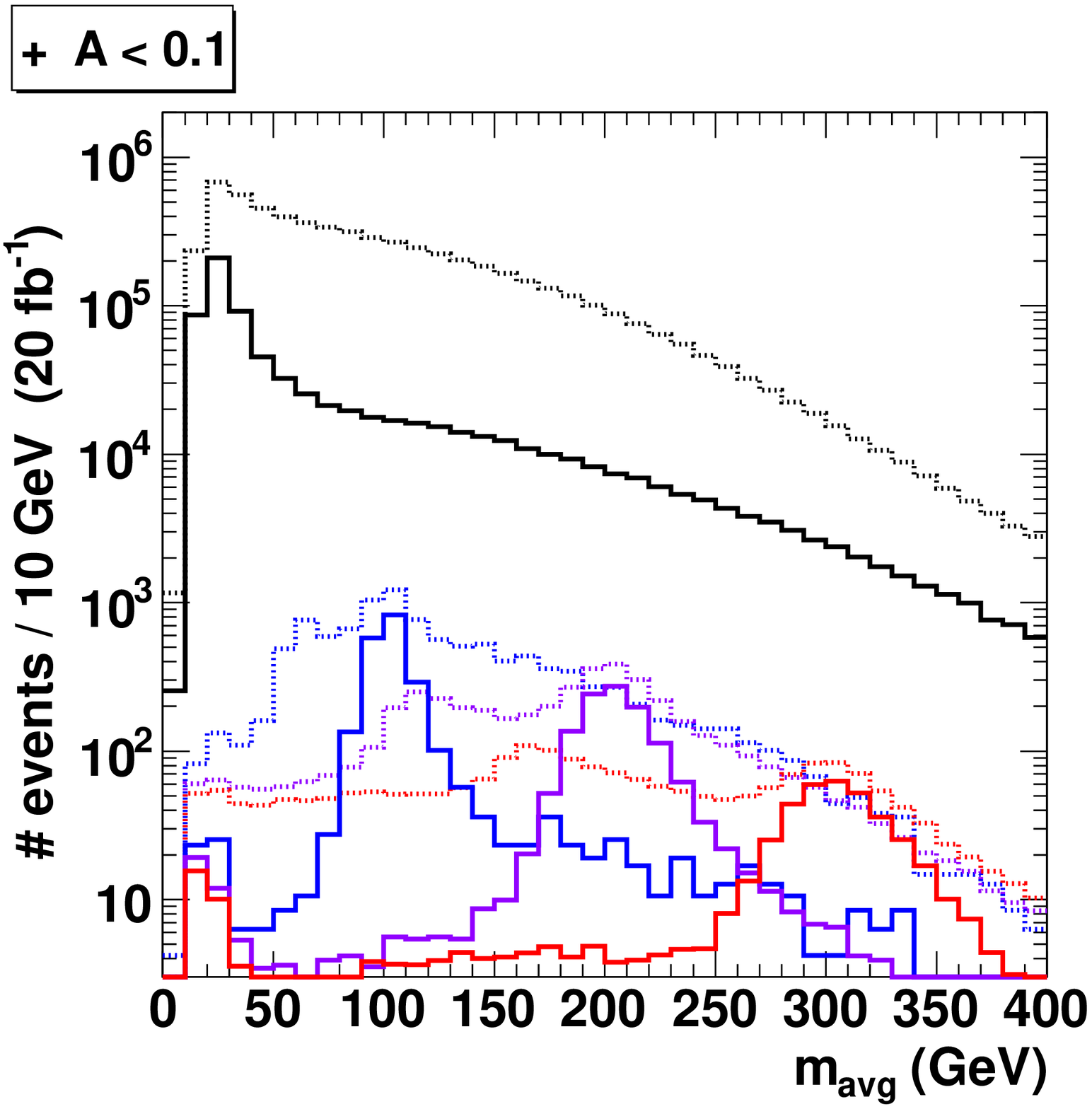}
\includegraphics[width=0.44\textwidth]{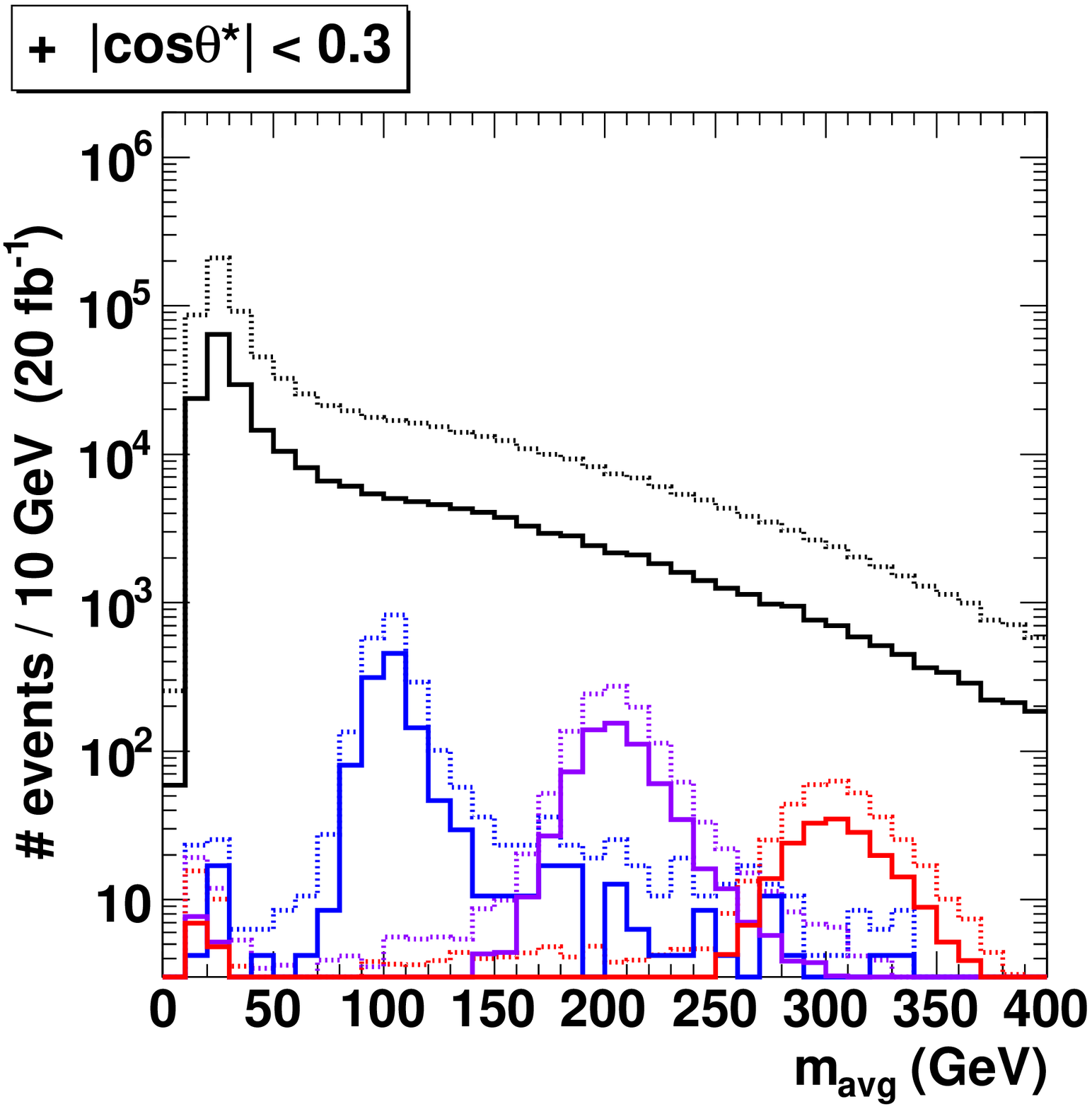}
\includegraphics[width=0.44\textwidth]{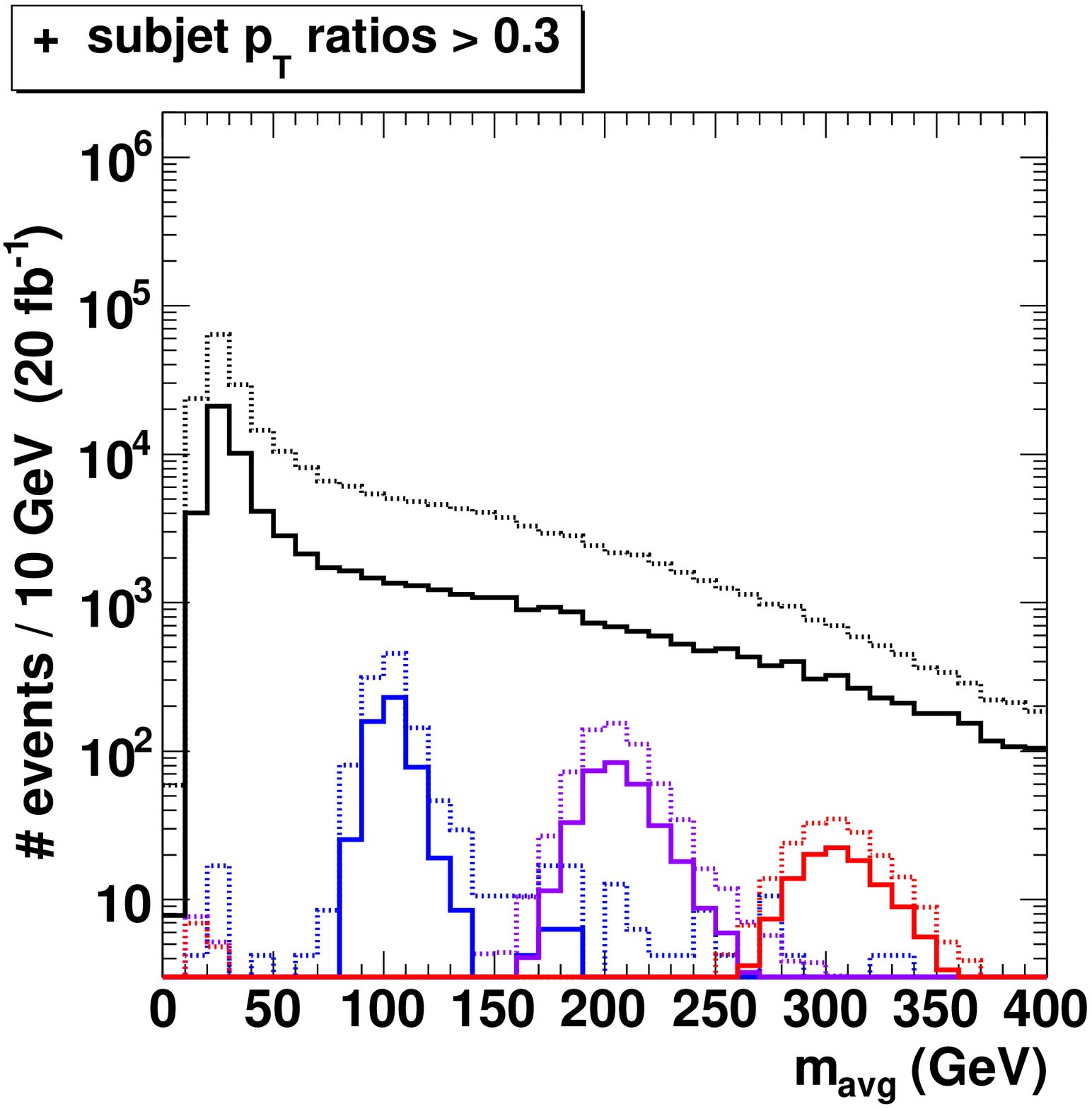}
\caption{The effects of our cuts on the spectrum of $m_{\rm avg} \equiv (m_1+m_2)/2$, defined on declustered fat-jets.  From left-to-right, top-to-bottom, cuts are added sequentially.  The effect of the preceding cut is shown with dotted histograms for comparison.  Background is matched QCD (black), and example stop models are 100~GeV (blue), 200~GeV (purple), and 300~GeV (red).}
\label{fig:cutFlow}
\end{center}
\end{figure*}

Figure~\ref{fig:cutFlow} shows the effect of the cut-flow on the $m_{\rm avg}$ spectra of continuum QCD and some example signal mass points.  There are several features that are worth noting.  First, the ubiquitous turn-on peak is present, but has been pushed down to the $O$(10~GeV) scale, far away from our signals.  This can easily be understood from the interplay between our $H_T$ cut, the 10\% relative-$p_T$ requirement in the declustering, and the size of our calorimeter cells.  (Note that the lowest-$p_T$ subjets that we work with are roughly 40~GeV.)  For the signal, before the $A < 0.1$ cut, it is possible to see three distinct mass features:  the turn-on, the true stop-mass peak, and an intermediate peak near half of the stop mass.  The last feature arises from events where one stop is correctly reconstructed, but one of the quarks was lost for the other stop.  After the $A$ cut, the stop peaks become very clear.  The additional cuts both reduce the overall size of the background and further tighten the signal peaks.  Throughout the entire set of cuts, the QCD background in the vicinity of the signals stays ``featureless.''  The final signal efficiencies relative to the inclusive pair production rates are $5\times10^{-5}$ for 100~GeV and $4\times10^{-3}$ for 300~GeV.  For comparison, stop pair acceptances for the standard 4-jet searches are usually at the $10^{-3}$--$10^{-2}$ level.

\begin{figure*}[ht!]
\begin{center}
\includegraphics[width=0.44\textwidth]{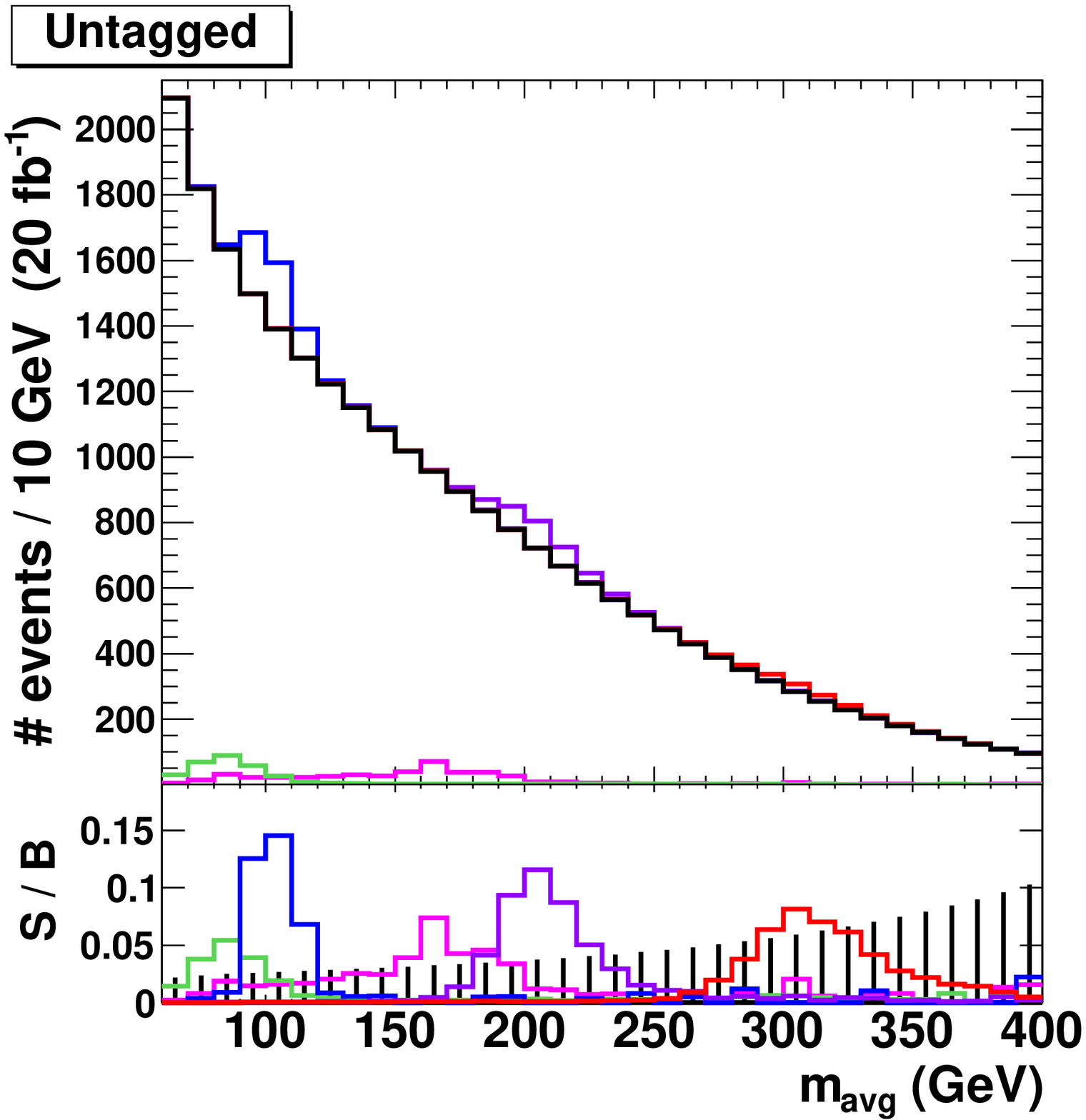}
\includegraphics[width=0.44\textwidth]{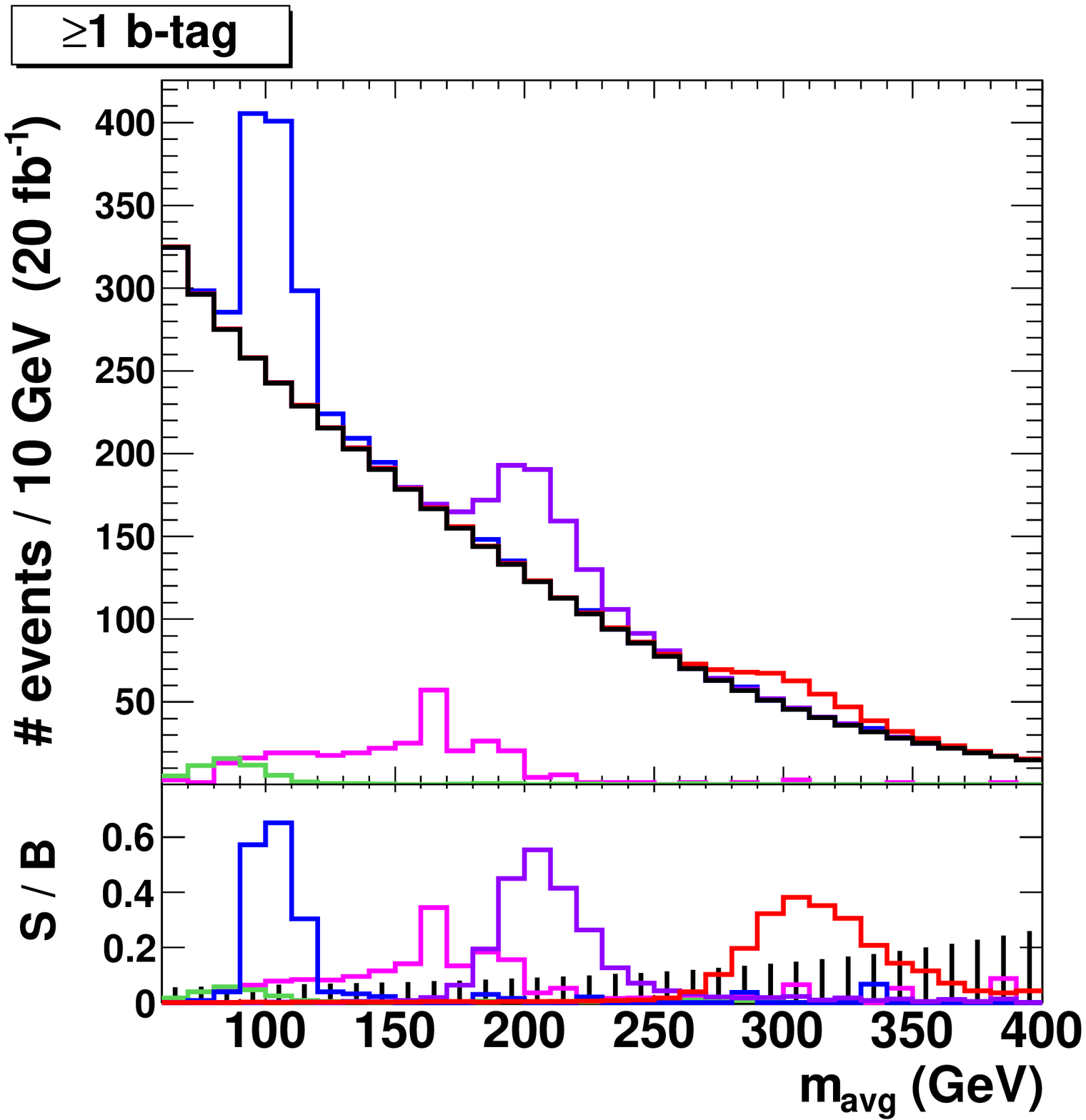}
\caption{Final spectra of $m_{\rm avg}$ after all cuts, for an untagged analysis (left) and a $b$-tagged analysis assuming $\mbox{BR}(\tilde t \to \bar b \bar d / \bar b \bar s) \simeq 100\%$ and tagging/mistagging rates as described in the text (right).  Displayed backgrounds include matched QCD (black), $t\bar t$ (pink), and $W$+jets (green).  The matched QCD histogram has been smoothed from the Monte Carlo data, as described in the next section.  Displayed example stop models, stacked onto the QCD background, include 100~GeV (blue), 200~GeV (purple), and 300~GeV (red).  The lower panels show the $S/B$ ratio relative to QCD, and the bin-by-bin fractional statistical errors on the QCD background expected for the 2012 LHC run.  (Note the changes in vertical axes between untagged and tagged.)}
\label{fig:mavg}
\end{center}
\end{figure*}

We show how the signal peaks appear on top of the continuum QCD background in linear scale in the left panel of Fig.~\ref{fig:mavg}, and give an indication of $S/B$ relative to the QCD background's bin-by-bin statistical errors.  It is clear already from this simple plot that the 100~GeV stop should be visible with high significance, and that exclusion reach should extend beyond 200~GeV.  We also show on this plot the largest subdominant background contributions, namely $t\bar t$ and $W$+jets.  (Other backgrounds such as diboson and single-top can be shown to be far smaller.)  We will not directly address the impact of these backgrounds on a search, but they are clearly important.  This is especially true for $t\bar t$, which contributes a broad bump peaked near 175~GeV, and at a size only $O(1)$ smaller than our stop signal.\footnote{The fact that $t\bar t$ is not a larger contribution is perhaps somewhat surprising, given that for $m_{\tilde t} \simeq m_t$, the inclusive $t\bar t$ cross section is about six times larger than $\tilde t \tilde t^*$.  About half of this factor comes from the $t\bar t$ all-hadronic branching fraction, since only all-hadronic events are efficient at passing the $H_T$ cut and subsequent substructure cuts.  It is also important to realize that for high-$p_T$ central production, the difference in cross sections is not as big.  (Asymptotically, the factor of six reduces to a factor of two.)  Finally, the large fraction of partial reconstructions with two-body substructure significantly broadens the top peak shape.}  However, the multibody structure of this background is under much better theoretical control than pure QCD, and its normalization could be extracted in the highly orthogonal semileptonic channel.  We therefore anticipate that it could be systematically subtracted or accounted for in a constrained fit.  Indeed, it can even serve as a useful calibration peak.  If it is necessary to further suppress $t\bar t$, it might be possible to do so with supplementary substructure cuts that can pick out and reject 3-body features, without highly rescultpting the continuum QCD.  (E.g., N-subjettiness~\cite{Thaler:2010tr} observables or the dimensionless variables of the HEPTopTagger~\cite{Plehn:2010st} would be appropriate to study.)  Regardless, some degradation of sensitivity in the vicinity of $m_t$ should be expected in reality.

If the RPV coupling obeys MFV, then almost every stop decay will contain a $b$-quark.  It therefore becomes possible to exploit a $b$-tagged analysis.  We show in the right panel of Fig.~\ref{fig:mavg} the $m_{\rm avg}$ spectra after demanding that at least one of the four subjets is tagged, assuming flat ($b$, $c$, $q/g$) tag rates of (60\%, 10\%, 2\%).  The $S/B$ (and $S/\sqrt{B}$) improves dramatically, as does the relative contribution of $t\bar t$ to the background budget.  Exclusion reach to above 300~GeV already appears highly likely.

These distributions set the stage for our data analysis in the next section.


\section{Search Strategies and Sensitivity Estimates}
\label{sec:discovery}

Extraction of a bump on top of a smoothly-falling background spectrum is a classic problem that has appeared many times already at the LHC.  Probably the most famous recent application is the observation of the Higgs resonance in the continuum diphoton spectrum~\cite{ATLAS-2013-012,CMS-PAS-HIG-13-016}, but this strategy has also been applied by CMS in its paired dijet resonance searches~\cite{CMS:coloron,Chatrchyan:2013izb}.  Relying solely on the assumption that the background is ``featureless,'' the observed spectrum can be fit to a parametrized function with or without the addition of a signal bump.  The parametrization used by CMS, which we also use here, is a four-parameter function of the form~\cite{Chatrchyan:2013izb}
\beq
\frac{dP}{d m_{\rm avg}} \,=\, p_0\times\frac{(1-m_{\rm avg}/\sqrt{s})^{p_1}}{(m_{\rm avg}/\sqrt{s})^{p_2+p_3\log(m_{\rm avg}/\sqrt{s})}}\,,  \label{eq:shape}
\eeq
where $p_0$, $p_1$, $p_2$, $p_3$ are free parameters and $\sqrt{s}$ is the proton-proton center-of-mass energy (8~TeV for 2012).\footnote{It may be possible to develop parametrizations that more directly incorporate the analytic structure of QCD (see~\cite{Dasgupta:2013ihk,Dasgupta:2013via}), though we have not explored this possibility.}

There are also many other ways to directly estimate the QCD $m_{\rm avg}$ spectrum from the data, using control regions.  The common ABCD method was used by ATLAS in its own paired dijet resonance searches~\cite{Aad:2011yh,Atlas:coloron}.  This method requires defining sideband cuts in two variables.  The nominal cuts define region~A, and the other three choices of 2D cuts (nominal+sideband, sideband+nominal, and sideband+sideband) define three signal-depleted regions B, C, and D.  Assuming small correlations between the two variables (an assumption that must be justified in Monte Carlo studies), the region-A background spectrum can be derived by taking the bin-by-bin ratios of counts B$\times$C/D.  For its ABCD-based search, ATLAS uses the variables $A$ and $|\cos\theta^*|$ to define its four regions.  We run our own version of this search, taking sideband cuts $A = [0.1,0.4]$ and $|\cos\theta^*| = [0.3,0.8]$.

In addition to the shape-fit and ABCD methods, we also explore two supplemental techniques which may further improve statistical power.

In the first of these, we run a simultaneous shape fit with Eq.~(\ref{eq:shape}) over our nominal signal region and a nearby asymmetry sideband region $A = [0.1,0.2]$.  This increases the data statistics available to the fit by $O(1)$, as well as folding in more discriminating characteristics between the stop signal and QCD.  The method is based on the assumption that the QCD spectrum changes very slowly as a function of $A$ when $A \ll 0$, whereas the signal is strongly peaked in $A$ for $m_{\rm avg} \simeq m_{\tilde t}$.  We observe exactly this behavior in our simulations.  To implement the simultaneous fit, we set the QCD fit parameters identical between the two $A$ regions.  We can then add into the fit the signal shapes and relative normalizations appropriate to each individual region.  This simple two-region fit might also be improved by more finely subdividing in $A$, and thereby more fully exploiting the nontrivial shape of the signal over this variable.  Equivalently, this can be viewed as a full 2D fit over the small-$A$ region of the $(m_1,m_2)$ plane.

The final method that we study (inspired in part by~\cite{Hedri:2013pvl}) is based on the assumption that the two fat-jet masses can be considered approximately uncorrelated in background events.  If this assumption holds, then the $m_{\rm avg}$ spectrum in the $A \to 0$ limit can be directly predicted from the spectra of individual fat-jets.  Given a single-jet probability distribution $dP/dm_{\rm jet}$ (where $m_{\rm jet}$ represents the subjet-pair mass), we get 
\beq
\left.\frac{dP}{dm_{\rm avg}}\right|_{A < A_{\rm cut} \ll 1} \,\sim\; \left[\frac{dP}{dm_{\rm jet}}(m_{\rm avg})\right]^2\times m_{\rm avg}.
\eeq
The major advantage of this method is that the statistics available for measuring $dP/dm_{\rm jet}$ are enormous, since no tight cuts should be placed on the event-by-event jet-mass asymmetry.  However, the uncorrelated assumption needs to be carefully studied in simulation.  (It may also be tested to some extent in data, by inverting some of our final cuts.)  To keep the kinematics similar to our final signal region, we measure the $dP/dm_{\rm jet}$ spectrum in a control region with only the $H_T$ and $|\cos\theta^*|$ cuts in place.  We pick a random jet amongst the leading two, and further demand the $\min[p_T(a),p_T(b)]/\max[p_T(a),p_T(b)]$ cut for only that jet.  (The signal contamination in this single-jet control sample is negligibly small.)  Once the $dP/dm_{\rm avg}$ spectrum estimate is derived, we use it as a template for a one-parameter normalization fit to our signal-region spectrum, with or without a signal bump added.  In practice, we approximate $dP/dm_{\rm jet}$ with a very finely-binned spectrum measured from our ``data,'' apply the above transformation, and then integrate back to our nominal $m_{\rm avg}$ binning (10~GeV).

\begin{figure*}[t!]
\begin{center}
\includegraphics[width=0.44\textwidth]{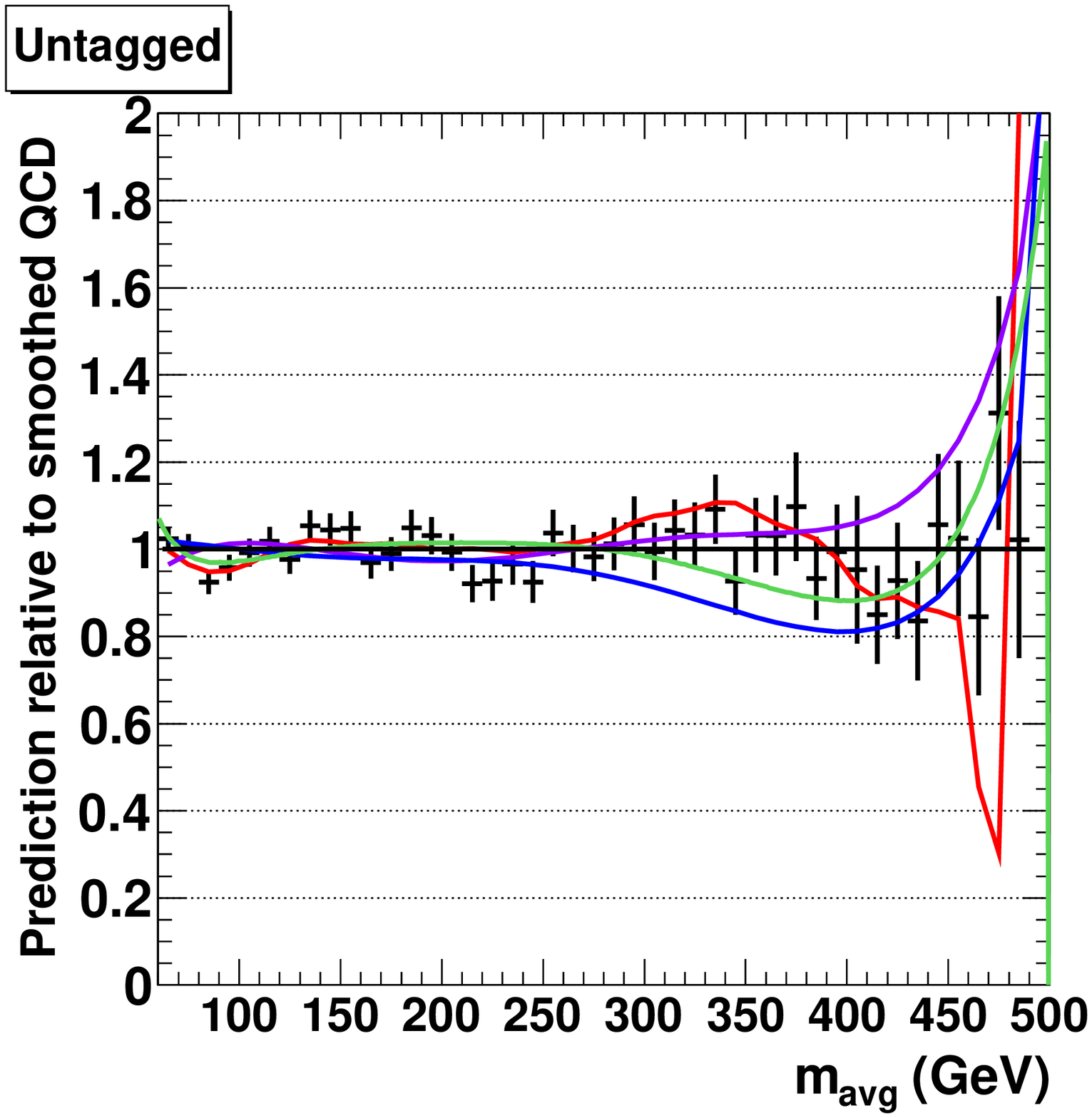}
\includegraphics[width=0.44\textwidth]{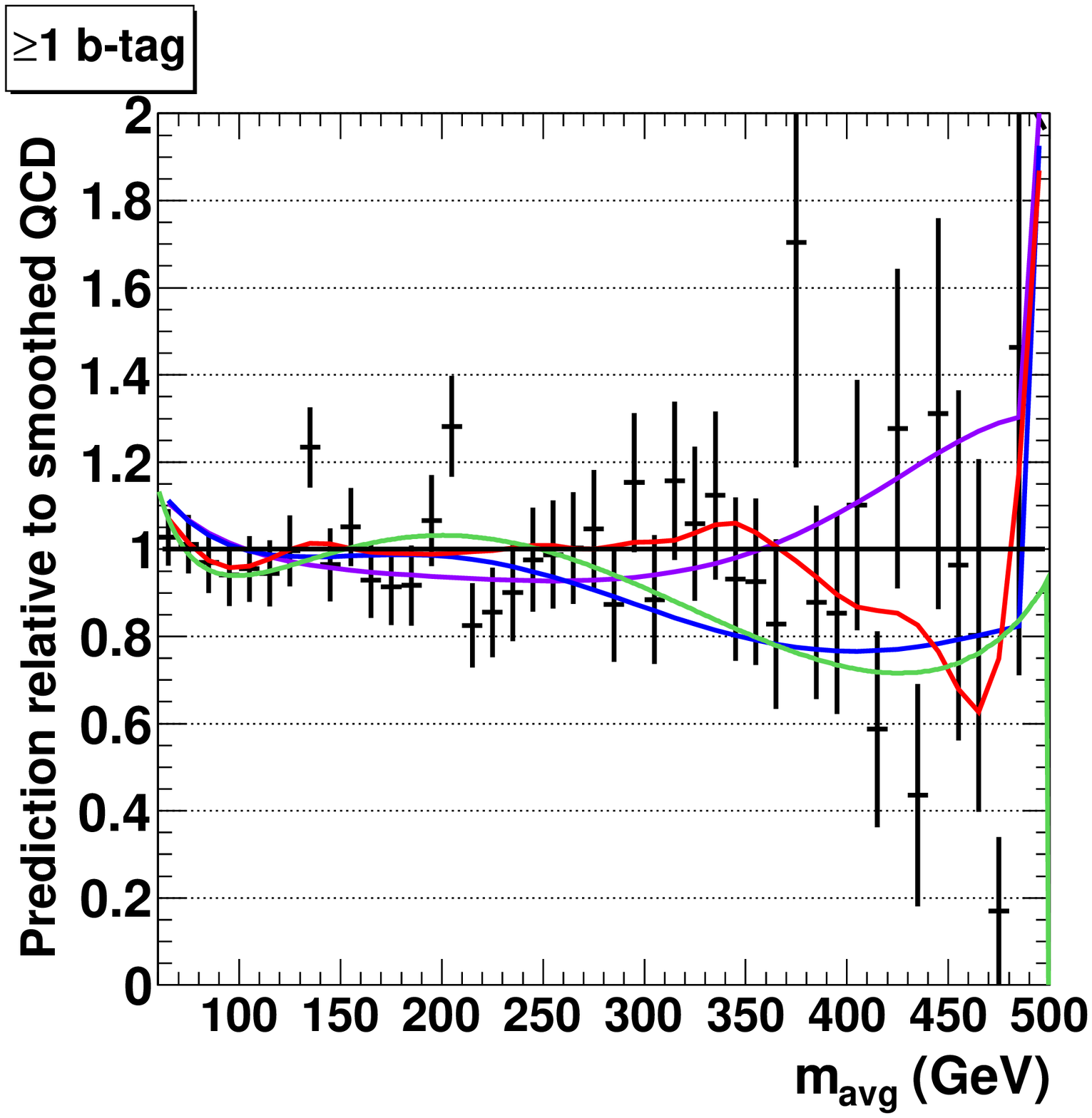}
\caption{Relative agreement between different methods of estimating the continuum QCD $m_{\rm avg}$ spectrum, for untagged (left) and $b$-tagged (right) analyses.  Here, ``1'' is defined as the prediction from our Monte Carlo data, smoothed with the exponential of a fifth-degree polynomial.  The black histogram is the original Monte Carlo data, with error bars estimated from the quadrature-sum of event weights.  The four curves correspond to four estimation methods:  shape fit applied to the original MC data (green), ABCD (purple), asymmetry-sideband (blue), and single-jet template (red).}
\label{fig:all_estimators}
\end{center}
\end{figure*}

Since our four search methods (shape fit, ABCD, asymmetry-sideband, and single-jet template) all rely on assumptions about the behavior of the QCD background, we can first get some sense of the validity of those assumptions.  To do so, we compare the signal-region QCD spectrum to the predictions of the four methods, without signal.  Here and below, we focus on the mass range $m_{\rm avg} = [60,500]$~GeV.  Also, due to the finite statistics of our QCD Monte Carlo sample, which is actually comparable to a 2012-like data set, we smooth out the spectra in the signal and control regions by fitting to the exponential of a fifth-degree polynomial.  These fits generally have high goodness-of-fit probability.\footnote{A similar parametrization could also be considered for the shape fit search method, or alternatively we could have smoothed our Monte Carlo data using the functional form of Eq.~(\ref{eq:shape}).  In fact, we have found that the latter does indeed furnish a high-probability fit.  However, we have explicitly chosen different functions for smoothing the raw Monte Carlo and for fitting the derived pseudodata, in order to help prevent spuriously good pseudodata fits.}  (We do not apply smoothing to the single-jet spectrum, which has much higher statistics.)  We show the results of the comparison in Fig.~\ref{fig:all_estimators}, including the original Monte Carlo data with its statistical errors.  For $m_{\rm avg} \ltap 300$~GeV, the agreement is generally better than 5\% (10\% with the smaller-statistics $b$-tagged sample).  At higher masses, more pronounced disagreements develop, but their significance is likely not large given the growing error bars.  This is also entering the $m_{avg}$ region where the subjets are nearly separated by $\Delta R=1.5$ and the spectrum is turning off, in which case we might not be surprised to find more sensitivity to the different control region cuts used in the estimation procedures.  Needless to say, a higher-statistics simulation would be useful here to better gauge the level of agreement at both higher and lower masses.  Still, given the encouraging agreement over the mass range in which we will shortly find our best sensitivity, we can proceed with our analyses without fear that their underlying assumptions are grossly invalid, at least for matched QCD.\footnote{We have also rerun the untagged comparison on a larger background sample generated wholly within {\tt Pythia8}, based on showered $2\to2$ production without matching.  The shape fit and ABCD methods continue to improve across the full mass range, generally agreeing with the signal region spectrum to better than 2\% for $m_{\rm avg} < 350$~GeV (and still consistent within MC errors).  The asymmetry-sideband agreement exhibits a -20\% dip above 300~GeV, similar to the matched case, but with much higher significance.  Interestingly, the single-jet template also develops a broad +20\% discrepancy above 300~GeV, which appears to be more severe than for the matched case.  Again, these features occur in regions where our sensitivity is in any case becoming poor, but they would need to be more carefully investigated in a fully realistic search.}

We now apply the four data-driven methods to search for the stop signal in our simulations, using the matched QCD background.  For all four methods, we search for a signal using $\chi^2$ differences as a discriminator.  In the case of the fits, we take the $\chi^2$ difference between background-only and signal+background fits, using $\sqrt{N}$ error bars for the observed bin counts.  For the ABCD method, which directly predicts the background spectrum with no free parameters, we construct $\chi^2$ bin-by-bin by combining in quadrature the statistical errors of the observed spectrum and the predicted spectrum.  (The latter is itself derived by simple propagation-of-errors from the B, C, and D bin counts.)  In all cases, the signal strength and shape are fixed, and systematic errors are not assessed.  Our results should therefore be indicative of what can be accomplished in the limit of small systematics.\footnote{Systematic errors affecting the signal shape and normalization will include the jet/subjet energy scales and resolution uncertainties, and probably PDF and scale errors for the high-$p_T$ cross section.  Possible systematics affecting our different data-driven estimates of the QCD spectrum would need to be investigated with higher-resolution Monte Carlo simulations with a full detector mock-up and/or in control data.}

We use our (smoothed) background simulations and signal+background simulations as the basis of a large number of pseudoexperiments corresponding to 20~fb$^{-1}$ at 8~TeV.  We run our various search strategies on these pseudoexperiments to build up $\Delta\chi^2$ distributions.  We find these distributions to be highly Gaussian, at least out to about three standard deviations.  To better parametrize their separation, we therefore fit the $\Delta\chi^2$ distributions to Gaussians.  From these we can derive a ``median discovery significance'' and a ``median exclusion significance.''  These are the distance between the medians of the two $\Delta\chi^2$ distributions, as measured in units of the background-only pseudoexperiments' $\sigma$ (for discovery) or the signal+background pseudoexperiments' $\sigma$ (for exclusion).  We can set a benchmark for discovery at 5$\sigma_{B}$ separation, not accounting for the look-elsewhere effect.  We can set a benchmark for exclusion at 2$\sigma_{S+B}$, which is practically equivalent to the usual 95\% $CL_S$ criterion.  For exclusion, we also consider fluctuations at $\pm1\sigma_B$ about the background-only median, and recompute the exclusion level in $\sigma_{S+B}$ units.  

\begin{figure*}[t!]
\begin{center}
\includegraphics[width=0.44\textwidth]{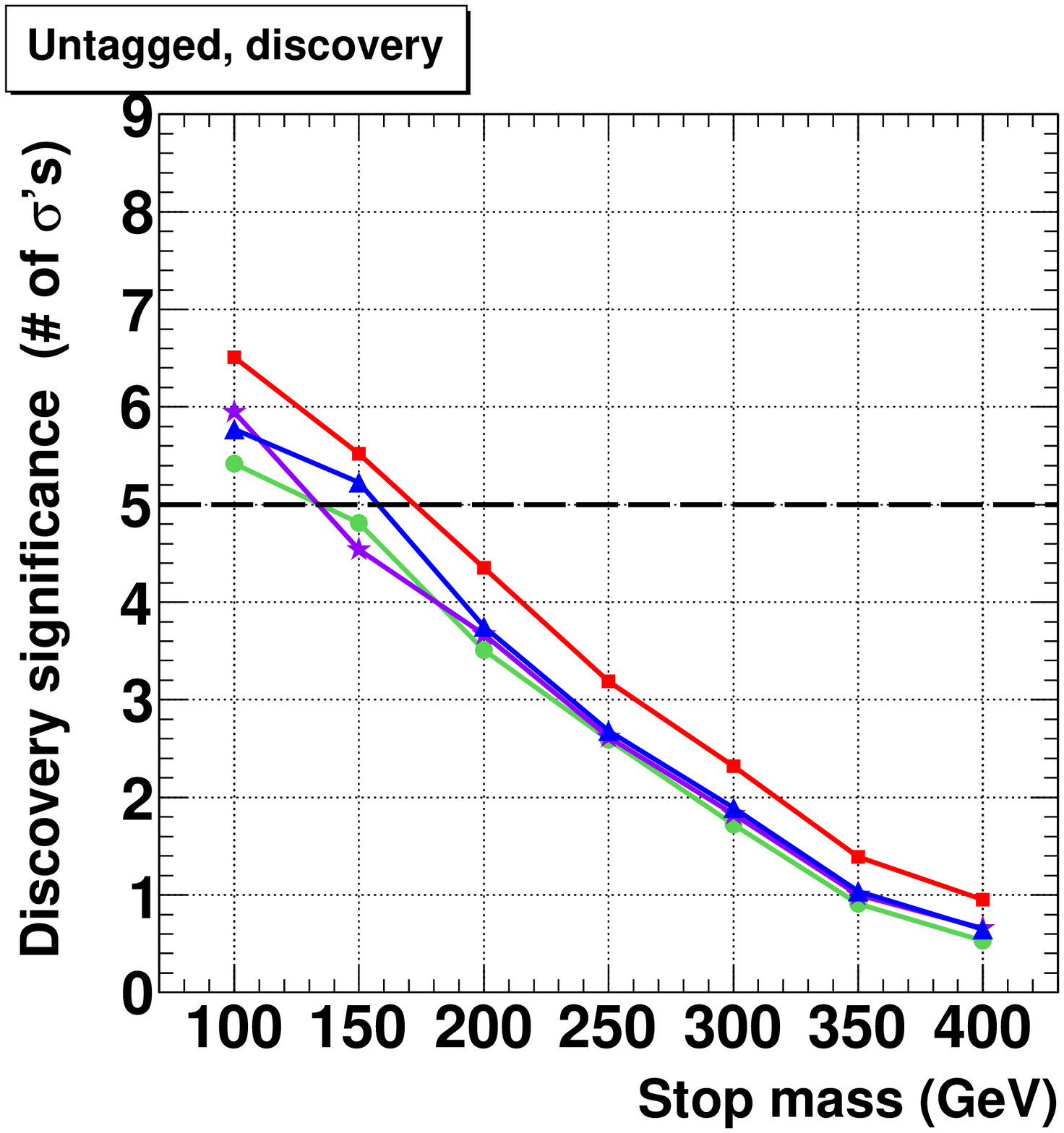}
\includegraphics[width=0.44\textwidth]{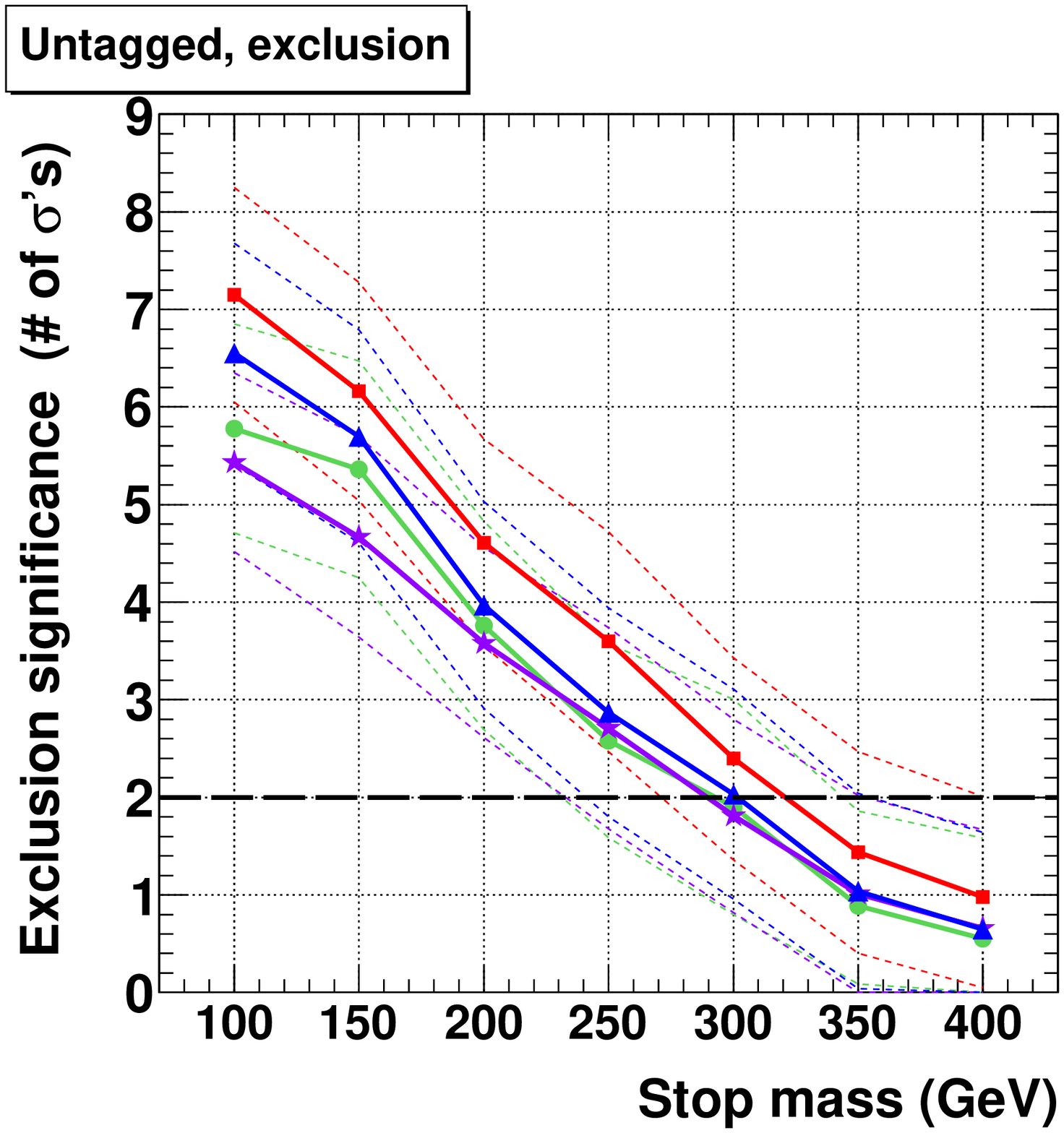}
\caption{Results of our untagged search estimates for 20~fb$^{-1}$ at LHC8, displaying median discovery significance (left) and exclusion significance (right).  The four curves correspond to four data-driven QCD background estimation methods:  shape fit (green), ABCD (purple), asymmetry-sideband (blue), and single-jet template (red).  In the exclusion significance plot, we also indicate the $\pm1\sigma$ variation expected due to background statistical fluctuations.}
\label{fig:results_untagged}
\end{center}
\end{figure*}

\begin{figure*}[t!]
\begin{center}
\includegraphics[width=0.44\textwidth]{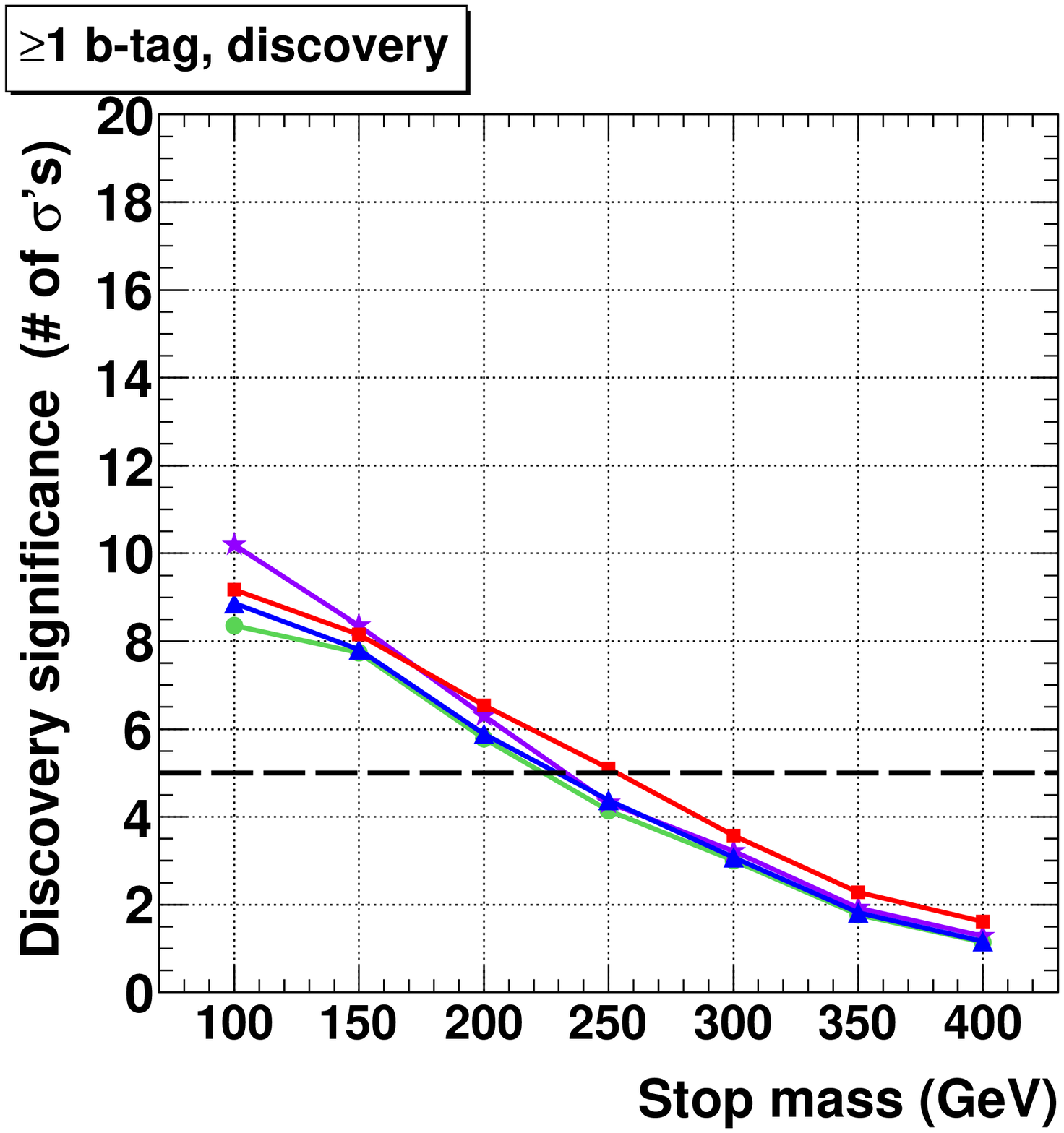}
\includegraphics[width=0.44\textwidth]{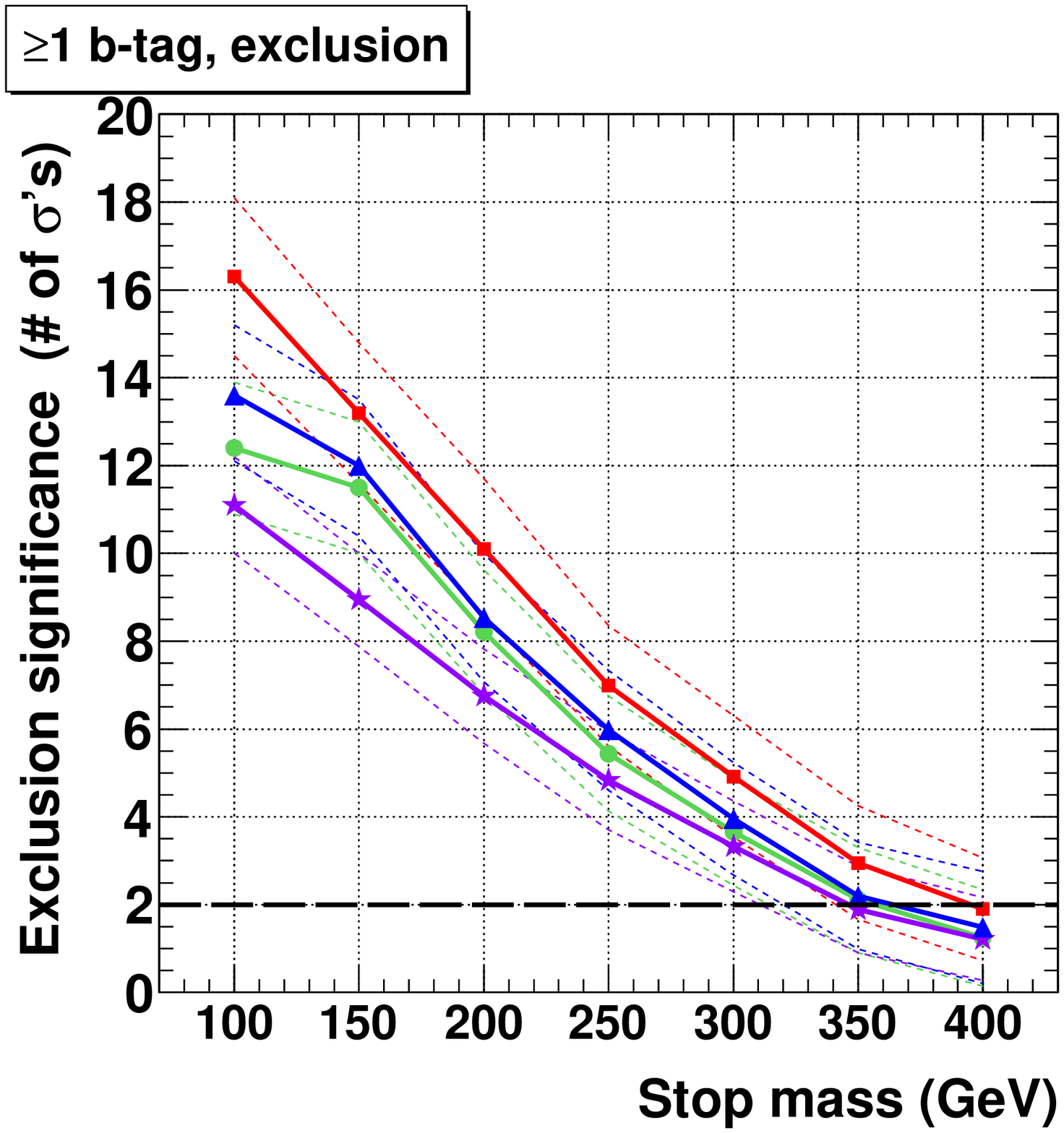}
\caption{Results of our $b$-tagged search estimates for 20~fb$^{-1}$ at LHC8, displaying median discovery significance (left) and exclusion significance (right).  The four curves correspond to four data-driven QCD background estimation methods:  shape fit (green), ABCD (purple), asymmetry-sideband (blue), and single-jet template (red).  In the exclusion significance plot, we also indicate the $\pm1\sigma$ variation expected due to background statistical fluctuations.}
\label{fig:results_tagged}
\end{center}
\end{figure*}

Figs.~\ref{fig:results_untagged} and~\ref{fig:results_tagged} show the final results for untagged and $b$-tagged analyses, respectively. 
It is clear that all four search methods perform comparably, but that the single-jet template method tends to edge out the other three, and that the asymmetry-sideband method offers a small but consistent improvement over the simple shape fit.  (In fact, for exclusion significance, the single-jet template method gives results very close to what would be inferred with a naive $S/\sqrt{B}$ analysis with optimized mass windows.)  The similarity of the results is encouraging, and suggests that experimentalists will have many alternative choices for performing cross-checks of a tentative signal, or as fall-back options if any of these data-driven methods turns out to be unreliable.  From Fig.~\ref{fig:results_untagged}, which shows the untagged analysis, we see that stops less than about 175~GeV could be discovered, and stops less than about 320~GeV could be excluded.  For the $b$-tagged analysis in Fig.~\ref{fig:results_tagged}, masses below 250~GeV are discoverable, and exclusion sensitivity extends to nearly 400~GeV.  We note that this analysis was run without re-optimization of our cuts, so it might be possible to construct an even more sensitive search.  It may also be possible to make even further gains by considering a double-$b$-tagged search.

Looking ahead, we have also run versions of these analyses on 14~TeV simulations, assuming 300~fb$^{-1}$ luminosity, and for simplicity neglecting pileup.  Here, we have used a summed-jet $H_T$ cut of 1600~GeV, which keeps the rate approximately the same as the 900~GeV threshold under 2012 conditions (assuming quadrupled instantaneous luminosity at 14~TeV).  The 100~GeV untagged signal remains visible, with statistical significance slightly better than our 2012 estimate, though with approximately 2--3 times smaller $S/B$.  The discoverable range expands up to about 500~GeV, and masses of 200--300~GeV would be visible at the 10$\sigma$-level.  Exclusion should extend up to 650~GeV.  This last finding is comparable to that of the recent Snowmass~2013 report~\cite{Duggan:2013yna}, which uses traditional jet reconstruction methods and a highly approximate background estimate.  However, that search assumes 2012-like jet $p_T$ cuts, and even then is limited to the mass range above 300~GeV.  By contrast, in our jet substructure version of the search there is practically no low-mass cutoff on the search range, with masses from 100~GeV to $O$(TeV) covered by a single analysis strategy.


\section{Conclusions}
\label{sec:conclusions}

In this paper, we have addressed what has been believed to be one of the most difficult supersymmetry signatures at hadron colliders, and demonstrated that it may nonetheless be made highly visible using the tools of jet substructure.  Besides serving as a crucial supplement to the LHC's broad-based program for testing naturalness, this result, if reproducible in a realistic analysis on actual LHC data, will serve as a benchmark for fully jetty searches.  The implications extend well beyond just RPV supersymmetry.  Thus far, multijet searches at the LHC have successfully placed constraints on the pair-production a variety of colored objects, whose prompt decays contain neither leptons, neutrinos, nor other invisible particles~\cite{ATLAS-2013-091,ATLAS:2012dp,CMS-PAS-EXO-12-049,Chatrchyan:2012uxa,CMS:coloron,Chatrchyan:2013izb,Aad:2011yh,Atlas:coloron}.  However, compared to the LSP stop with baryonic RPV, which is a color-triplet scalar undergoing a two-body decay, these searches have always relied on more color, more spin, more flavor, and/or more final-state partons.  I.e., they can exploit much higher cross sections and/or more complicated event topologies.  With the search that we are proposing, we have finally ``hit bottom'' on colored particle pair-production and decay.  {\it Any} model with a color-triplet ``diquark'' scalar 
can be searched for, whether it is connected to naturalness or not, and independently of the flavor structure of its decay (see e.g.~\cite{Ma:1998pi,DelNobile:2009st,Giudice:2011ak,Vecchi:2011ab}).\footnote{The other choice of diquark color representation, namely the color-sextet, has not been explicitly searched for, but its cross section is actually larger than the color-octet.  We therefore expect that the limits on color-sextet scalars with prompt two-body decays are more constraining than those for color-octets~\cite{Aad:2011yh,Atlas:coloron}.}

In terms of concrete performance, we have found that jet substructure at the LHC can conclusively push beyond the 100~GeV threshold set by previous limits, leaving no gaps.  In fact, it should be possible to achieve discovery-level sensitivity at 100~GeV using 2012 data, demonstrating the LHC's far superior production cross sections and luminosity compared to earlier experiments.  Exclusion-level sensitivity extending up to 300~GeV or higher seems achievable.  These results can be obtained using a number of promising data-driven search techniques, and should be realistic if at least one of these techniques exhibits managable systematic errors.  Returning to the usual MFV-inspired assumption that most decays contain $b$-quarks, the option of an extremely powerful $b$-tagged search opens up.  This may provide roughly 10$\sigma$ discovery sensitivity at 100~GeV, and exclusions extending to nearly 400~GeV, without accounting for possible re-optimization of the analysis cuts or further gains from applying a double-$b$-tag.   The upcoming Run~II of the LHC should push exclusion to masses about twice as high, even while further tightening sensitivity at lower masses.

These findings have yet again illustrated the utility of viewing complicated hadronic activity through the lens of jet substructure.  This basic change in philosophy from canonical jet-based reconstruction (almost) frees us of the notion of a ``minimum distance'' between reconstructable hard partons, namely the jet radius, which was in fact a major botteleneck in the traditional 4-jet versions of this search.  Clearly with a more flexible viewpoint, the extremely high energies available at the LHC can be made to work for us, not against us, when attempting to search for $O$(100~GeV) or even $O$(10's~of~GeV) objects decaying into jets.  Indeed, we have found here that we are in a good position to test whether supersymmetry has exploited the limitations of our conventional analyses, and has been hiding in plain sight this whole time.


\acknowledgements{
We thank Jared Evans, Roberto Franceschini, and Keith Ulmer
for useful discussions and comments.  We thank Stefan Prestel for much help in using CKKW-L matching and for fixing a bug in {\tt Pythia8} {\tt v1.7.6}.  YB was supported by start-up funds from the Univ. of Wisconsin, Madison. 
AK was supported by NSF grant No.~PHY-0855591. BT was supported by DoE grant Nos.~DE-FG-02-91ER40676 and DE-FG-02-95ER40896, by NSF 
grant No.~PHY-0969510 (LHC Theory Initiative), and by the PITT PACC.  YB would also like to thank the Kavli Institute for Theoretical Physics, U.C. Santa Barbara, which is supported under Grant No.~NSF PHY11-25915.  AK is also grateful to the Galileo Galilei Institute in Florence where this work was partially done.
We also thank the Aspen Center for Physics, 
under NSF Grant No.~PHY-1066293, where this work was both initiated and completed.
}

\appendix

\section{Simulation Details}
\label{sec:simulations}

Our nominal stop pair simulations begin in {\tt MadGraph5} {\tt v1.4.7}~\cite{Alwall:2011uj} with a UFO simplified model~\cite{Degrande:2011ua} containing stops as the only BSM particle.  The simulations are matched up to one jet emission in production using the default {\tt PYTHIA6}~\cite{pythiamanual}.  The matching is of the $k_T$-MLM type, with a matching scale of 30~GeV.  There is a technical difficulty in such a simulation, in that {\tt PYTHIA6} does not recognize the color epsilon-tensor structure appearing in the stop's RPV two-body decay.  To bypass this difficulty, we first hadronize the stops, along with the rest of the event, using publicly-available R-hadronization code~\cite{rhadrons}.\footnote{When running with multi-parton interactions (i.e., underlying event), we found that this code runs much more seamlessly with the virtuality-ordered shower, and have used this for our simulations.}  We then simulate the stop-hadron decays by treating them as if they were $Z^*$s of the same mass, decaying to $b\bar d$ and showering/hadronizing independently of the rest of the event.\footnote{It is worth pointing out that, over a large range of acceptable parameters for the RPV couplings, the stops should indeed hadronize before decaying.  In any case, we expect that the order-of-operations of decay versus hadronization will only have a very minor impact on the final radiation pattern as seen by our procedures.}  This should build up a decay showering pattern essentially identical to what would have been obtained in {\tt Pythia8}~\cite{Desai:2011su}.

It is also possible to simulate stop pair events with RPV decays directly in {\tt Pythia8}, for which we use {\tt v1.7.6}~\cite{Sjostrand:2007gs}.  Our initial simulations indicated significant disagreements in the modeling of the production-stage radiation of high-$p_T(\tilde t)$ events relative to both matched {\tt MadGraph5} and {\tt Prospino~1.0}~\cite{Beenakker:1996ed}, and spurious $m_{\rm avg}$ tails in our final reconstructions.  Subsequently, we found that this behavior is driven by the default ``power shower'' treatment of events without light partons in the final state, and can be made to yield a good fit to the matrix element predictions by using shower damping ({\tt SpaceShower:pTdampMatch=1}).  (By contrast, we have found that the ``wimpy shower,'' as well as the virtuality-ordered and $p_T$-ordered showers of {\tt PYTHIA6}, all tend to underestimate the amount of production radiation.)  Using self-contained {\tt Pythia8} simulations with shower damping, and running through our complete analysis chain described below, we obtain very good agreement with the more complicated matched simulations.  Running matching within {\tt Pythia8} is also possible, but we have not explored this option for our signal generation.

Our nominal QCD continuum simulations are based on {\tt MadGraph5} matched up to four partons within {\tt Pythia8}, using the more theoretically-rigorous CKKW-L prescription~\cite{Lonnblad:2001iq}.\footnote{{\tt v1.7.6} of {\tt Pythia8} is affected by a bug that causes crashes for this matching.  We thank Stefan Prestel for providing us with a private fix.}  For our merging measure, we use the Durham-$k_T$ distance used internally by {\tt MadGraph5},\footnote{Beam-distances are computed as $k_{T,iB}^2 = p_{T,i}^2 + m_i^2$.  Inter-parton distances are computed as $k_{T,ij}^2 = \max(m_i^2,m_j^2) +  2\min(p_{T,i}^2,p_{T,j}^2)\times (\cosh\Delta\eta_{ij} - \cos\Delta\phi_{ij} )$.} and exploit the program's ability to produce multi-parton simulations with kinematic cuts defined in that space with threshold {\tt xqcut}.\footnote{To avoid double-counting of $\alpha_s$ reweightings, we have commented-out the relevant code in {\tt MadGraph5}.}  The measure only applies to partons that can realistically be viewed as merging within a QCD diagram according to quark flavor.  For example, two antiquarks would never be compared, nor would a $u$-quark and a $c$-antiquark.  We also forbid mergings with the beam if they would violate flavor, though this occurrence is very rare.  Our merging scale is set to 50~GeV for 8~TeV simulations, and 100~GeV for 14~TeV simulations.  (For the former, the $m_{\rm avg}$ range over which we run our stop searches is highly dominated by the hard 4-parton events.)

For comparison, we have also generated a set of unmatched QCD simulations wholly within {\tt Pythia8}, based on showered $2\to2$ production with the default (wimpy) shower.  These display a qualitatively similar $m_{\rm avg}$ spectrum to the matched sample, though with a less steep falloff, and $O(1)$ higher rate at $m_{\rm avg} \gtap 200$~GeV.  These differences are much less pronounced when comparing samples processed through ordinary jet reconstruction.


For our other backgrounds ($t\bar t$, $W$+jets, $t/s$-channel single-top, $tW$, diboson), we mainly relied on {\tt Pythia8}, though our $t\bar t$ simulation starts as a $2\to6$ decay chain processes in {\tt MadGraph5} in order to capture spin correlation effects.  We have also damped the shower for $t\bar t$, which is treated as a power shower by default.  Similar to stop pair production, comparison to the kinematics of undecayed matched samples shows good agreement.

All of our simulations are leading-order, and can be normalized to higher-precision calculations using K-factors.  For our matched stop samples, we find that a flat K-factor of $1.5$ corrects us to the NLO+NLL predictions~\cite{Beenakker:2010nq} for almost any mass.  We have also verified, by comparing with $p_T(\tilde t)$ spectra predicted by {\tt Prospino~1.0}, that the matched spectrum is in excellent agreement with the NLO spectrum.  Therefore we do not anticipate a substantially different K-factor for boosted stops.  For $t\bar t$, we use a K-factor of $1.8$.  For all of our other simulations, including matched QCD, we coarsely assume K-factors of $1.5$.  Our own comparisons of matched and unmatched QCD simulations to the data of~\cite{Chatrchyan:2013izb} suggest that this choice of K-factor may be conservatively large.

Downstream of the hadron-level simulation, we apply a simple detector model in the form of a $0.1\times0.1$ calorimeter grid in $\eta$--$\phi$ space.  We form our final $H_T$ trigger by first clustering the cells into $R=0.5$ anti-$k_T$ jets with {\tt FastJet3}~\cite{Cacciari:2005hq}, and demanding that the sum of $p_T$s of jets above 50~GeV exceeds 900~GeV (1600~GeV for 14~TeV simulations).  At this stage we neglect both jet energy measurement fluctuations and pileup effects.  The former should be only a few percent for such a large sum-over-energies, and the latter would be systematically accounted for in a realistic jet measurement.

While we do not concern ourselves with the effects of pileup on ordinary jet energy measurements, we do wish to make sure that our subsequent jet substructure methods are not adversely affected.  Therefore, on top of each hard event surviving the $H_T$ trigger, we superimpose an average of 20 min-bias events from {\tt Pythia8}, including both charged and neutral activity.  We then actively remove this pileup using a slightly modified form of trimming~\cite{Krohn:2009th}.  We cluster all calorimeter cells in the event into $R=0.2$ anti-$k_T$ jets, and discard the contents of these jets if their total $p_T$ falls below an absolute cutoff of 5~GeV.  Unlike canonical trimming, which works jet-by-jet and operates with a {\it relative} $p_T$ measure, we have targeted this method to subtract the contaminating energy density of pileup, at least in regions that do not overlap with hard activity.  We have found that this procedure successfully preserves our 100~GeV stop lineshape, which is otherwise shifted by 5~GeV and broadened due to the pileup.  The stop reconstruction rate is unchanged.  The final impact on the background is generally modest, though there is an $O(1)$ reduction of the high-$m_{\rm avg}$ tail, e.g. in the vicinity of 300~GeV.

Following this pre-trimming stage, we apply our main clustering into $R=1.5$ C/A fat-jets.  Only the leading two fat-jets in $p_T$ are considered, and these must both fall in the region $|\eta| < 2.5$.  The fat-jets are declustered into subjets as described in Section~\ref{sec:techniques}.  To obtain more realistic stop mass peaks, we smear the energies of the subjets as\footnote{More ideally, we would smear the energies of individual calorimeter cells, which would also allow us to simulate how imperfections in the energy measurements affect the declustering.  However, as we do not expect to be able to model this well, we still treat the energy measurements as perfect at that stage.  Still, our final analysis cut on the $p_T$ ratio of the subjets operates on smeared objects.}
\beq
\frac{\Delta E}{E} \,=\, \frac{5\;{\rm GeV}}{E} \,\oplus\, \frac{0.5\;{\rm GeV}^{1/2}}{\sqrt{E}} \,\oplus\, 0.05~.
\eeq

To cover scenarios where stops have large branching fractions to $b$-quarks, we also run a $b$-tagged analysis.  We assume that subjets can be tagged similar to ordinary jets, and that the tag/mistag efficiencies are fairly flat in our analysis range.  (Our subjet $p_T$'s typically vary between 100~GeV and 400~GeV.  See, e.g.,~\cite{CMS-PAS-BTV-13-001}.)  To perform flavor tagging, we keep track of bottom-hadrons and prompt charm-hadrons from the event record, and match them to the closest subjet within $\Delta R < 0.2$.  Each subjet's ``true'' flavor is then determined by the heaviest associated hadron.  We apply flat $b$-tagging efficiencies of 60\%, 10\%, and 2\% for bottom-flavored, charm-flavored, and unflavored subjets, respectively.


\section{Supplementary Results}
\label{sec:substructure}

This appendix contains three supplementary sets of results:  the $\Delta R$ distributions of subjets for signal events, a comparison of our nominal $R=1.5$ jet radius to $R=0.8$, and comparisons with the more standard BDRS declustering procedure.

\begin{figure*}[t!]
\begin{center}
\includegraphics[width=0.44\textwidth]{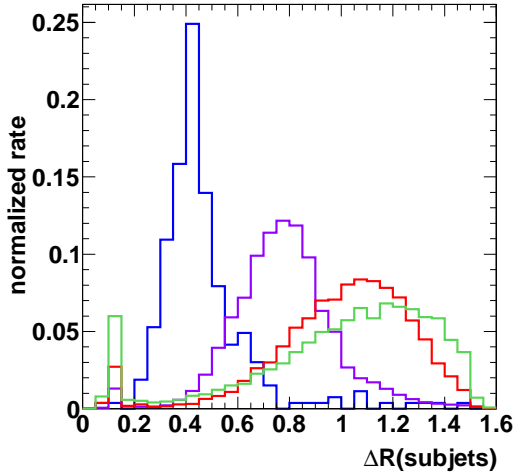}
\caption{The $\Delta R$ distributions of subjets within reconstructed fat-jets passing all analysis cuts, for stops of mass 100~GeV (blue), 200~GeV (purple), 300~GeV (red), and 400~GeV (green).  (Small spikes at $\Delta R = 0.1$ correspond to events where both stop-jets have been declustered down to our calorimeter model granularity, and would have $m_{\rm avg} \sim 10$~GeV.)}
\label{fig:DeltaRs}
\end{center}
\end{figure*}

Fig.~\ref{fig:DeltaRs} shows the $\Delta R$ distributions of subjets within stop-jets, for events passing our complete set of analysis cuts.  This plot makes it clear that for $m_{\tilde t} = 100$~GeV, a large fraction of stop decays would comfortably sit inside of a normal-sized LHC jet of $R=0.4$ or $R=0.5$.  It is also notable that, even though we choose a much larger fat-jet radius, very few stop decays are reconstructed with unphysically-large $\Delta R$.  In other words, our substructure procedures and analysis cuts adaptively find the ``correct'' $\Delta R$ scale for the signal.  For larger stop masses, the separation becomes large enough that an ordinary jet radius could resolve the decays.  But in our treatment this regime is continuously connected to the scenarios with $\Delta R < 0.4$, with no artificial threshold.  Finally, we can see that with our absolute and relative energy cuts, $m_{\tilde t} = 300$~GeV is about the largest mass that displays complete containment within $R=1.5$ fat-jets.  Still, a large fraction of $m_{\tilde t} = 400$~GeV decays remain contained, a signal which is important for the $b$-tagged version of the analysis.

\begin{figure*}[t!]
\begin{center}
\includegraphics[width=0.44\textwidth]{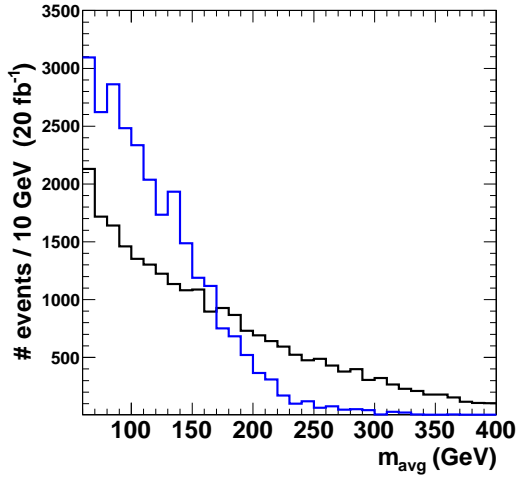}
\caption{Spectra in $m_{\rm avg}$ for matched QCD passing all cuts, reconstructed with the nominal $R=1.5$ fat-jets (black) or with $R=0.8$ (blue).}
\label{fig:R08comparison}
\end{center}
\end{figure*}

In Fig.~\ref{fig:R08comparison}, we compare the QCD continuum's $m_{\rm avg}$ spectrum with our nominal $R=1.5$ to an identical analysis with $R=0.8$.  It can be seen that, in the vicinity of $m_{\rm avg} = 100$~GeV, the background increases both in absolute rate and in steepness.  Essentially, the entire spectrum has been ``squashed'' by a factor of 2, since the overall mass scale is set by $R\times H_T$.  Performing the same analysis with the $m_{\tilde t} = 100$~GeV signal, the lineshape is practically unaltered, but the overall acceptance increases by 30\%.  This is because, with a narrower fat-jet, there are fewer cases where the declustering picks up a spurious ISR jet.  Still, the gain in $S/\sqrt{B}$ is marginal, and comes at a cost of slightly reduced $S/B$ in addition to a more difficult background shape.  Higher stop masses display significantly reduced efficiencies due to incomplete containment.

\begin{figure*}[t!]
\begin{center}
\includegraphics[width=0.44\textwidth]{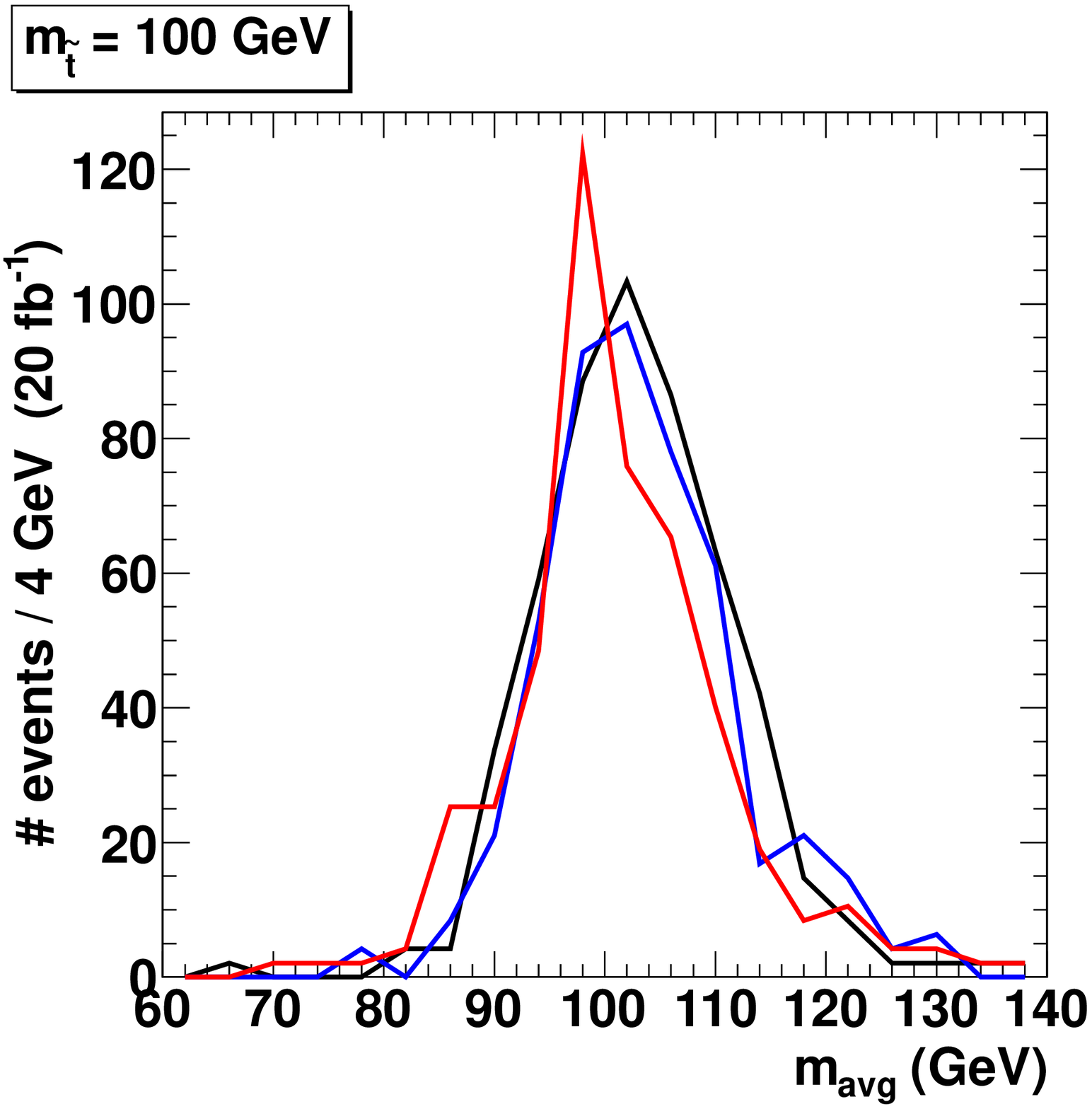}
\includegraphics[width=0.44\textwidth]{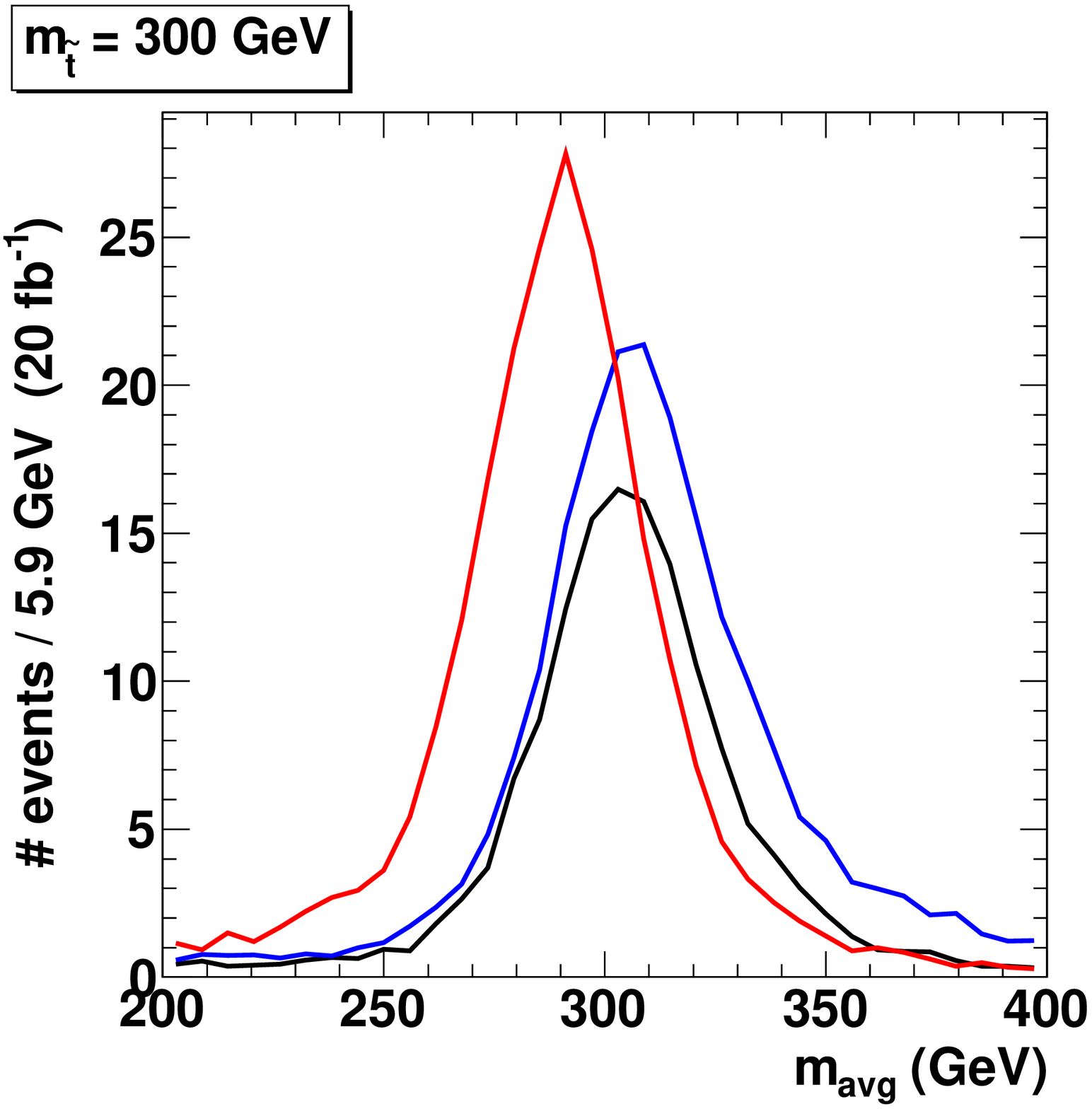}
\caption{Stop signal peak reconstructions for $m_{\tilde t} = 100$~GeV (left) and $m_{\tilde t} = 300$~GeV (right), using our nominal relative-$p_T$ declustering (black), full BDRS with filtering (red), and BDRS without filtering (blue).  (No pileup or trimming have been applied.)}
\label{fig:BDRScomparison_stop}
\end{center}
\end{figure*}

\begin{figure*}[ht!]
\begin{center}
\includegraphics[width=0.44\textwidth]{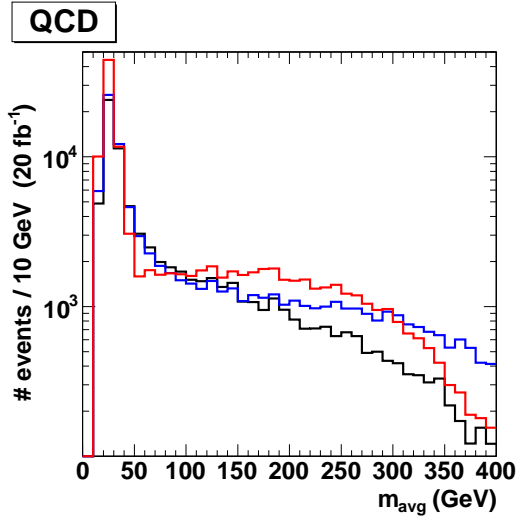}
\caption{Matched QCD reconstructions, using our nominal relative-$p_T$ declustering (black), full BDRS with filtering (red), and BDRS without filtering (blue).  (No pileup or trimming have been applied.)}
\label{fig:BDRScomparison_QCD}
\end{center}
\end{figure*}

Our nominal declustering procedure judges splittings based on the $p_T$'s of subjet candidates relative to the original fat-jet, and on their individual $m/p_T$ ratios.  This procedure is a direct descendant of the BDRS procedure of~\cite{Butterworth:2008iy}, which uses a somewhat different set of declustering criteria, and also applies an additional filtering step by reclustering the subjet constituents.  Within the kinematic regime of our analysis, the declustering stage of BDRS acts almost identically to our procedure without the $m/p_T$ requirement.\footnote{The BDRS mass-drop criterion is mostly redundant here, and the declustering is driven mainly by the momentum-asymmetry criterion.  See~\cite{Cui:2010km} for a detailed related study.}  With filtering, the two subjets are further refined into three, using the C/A algorithm with $R = \min[\Delta R({\rm subjets})/2,0.3]$.  We reform these back into two subjets, by clustering together the two that are closest in $\Delta R$.  This allows us to apply our final cut on the ratio of subjet $p_T$'s, which assumes 2-body substructure.  We also consider a form of BDRS without the filtering step.  Figs~\ref{fig:BDRScomparison_stop} and~\ref{fig:BDRScomparison_QCD} show a comparison of our nominal procedure with both filtered and unfiltered BDRS.  (This comparison is made without pileup or trimming.)  It is clear that the 100~GeV stop signal is fairly insensitive to the detailed procedure, but that the 300~GeV stop and QCD spectra can be highly reshaped, and that the overall rates in the vicinity of $m_{\rm avg} = 300$~GeV are increased by $O(1)$.  The filtered QCD spectrum is also flatter in the region between 100~GeV and 200~GeV, which could help a signal bump stand out more clearly there.  These last two points can be considered advantages.  However, filtering also accidentally introduces new mass scales from its minimum reclustering radius of $0.3$ for well-separated subjets, resulting in a bimodal QCD distribution with a local minimum near 50~GeV and a broad local maximum near 175~GeV.  And while the 300~GeV stop signal is enhanced, it sits on top of a background that is starting to sharply change shape.  Therefore, these advantages must be treated with caution.  Removing filtering, we still see some enhancement of the 300~GeV signal and its background, but the background shape becomes less biased.  The differences between unfiltered BDRS and our nominal declustering are dominated by the introduction of high-$m/p_T$ subjets for the former, and these are more likely to be contaminated by additional radiation.  We have also found that the detailed shape of the spectrum at high mass develops more sensitivity to changing analysis cuts to define control regions.  However, these differences relative our nominal procedure might be reduced with the use of trimming.


\onecolumngrid

\bibliography{lit}
\bibliographystyle{apsper}

\end{document}